\documentclass[preprint]{aastex}

\begin{document}

\title{Red Supergiants in the {\it JWST} Era. I: Near-IR Photometric Diagnostics}

\author{Emily M. Levesque}
\affil{Astronomy Department, Box 351580, University of Washington, Seattle, WA 98195, USA}

\shortauthors{Levesque} 
\shorttitle{RSGs with JWST: Near-IR Photometry}

\begin{abstract}
The Near Infrared Camera (NIRCam) on the James Webb Space Telescope ({\it JWST}) will be an incredibly powerful instrument for studying red supergiants (RSGs). The high luminosities and red peak wavelengths of these stars make them ideal targets for {\it JWST}/NIRCam. With effective photometric diagnostics in place, imaging RSG populations in multiple filters will make it possible to determine these stars' physical properties and, in cases where {\it JWST} pre-explosion imaging is available, to identify RSG supernova progenitors. This paper uses observed and model spectra of Galactic RSGs to simulate {\it JWST}/NIRCam near-IR photometry and colors, quantify and test potential diagnostics of effective temperature and bolometric magnitude, and present photometric techniques for separating background RSG and foreground dwarf populations. While results are presented for the full suite of near-IR filters, this work shows that (F070W-F200W) is the {\it JWST}/NIRCam color index most sensitive to effective temperature, F090W is the best band for determining bolometric magnitude, and the (F070W-F090W) vs. (F090W-F200W) color-color diagram can be used to separate foreground dwarf and background RSG samples. The combination of these three filters is recommended as the best suite of photometric observations to use when studying RSGs with {\it JWST}.
\end{abstract}

\maketitle

\section{Introduction}
Red supergiants (RSGs) are the He-fusing descendants of moderately massive ($\sim$10-25$M_{\odot}$) main sequence stars, the end result of a nearly horizontal evolution along the Hertzsprung-Russell (H-R) diagram. They are the coldest and largest members of the massive star population, sitting at the Hayashi limit (Hayashi \& Hoshi 1961) on the H-R diagram, and the most luminous K- and M-type stars in the universe, marking them as an important and observationally-valuable extreme in the evolution of massive stars.

Photometric criteria have been used to identify RSG populations at distances ranging from within our own Milky Way to well beyond the Local Group, spanning a broad range of host environments and metallicities. Populations of RSGs with well-determined physical properties - in particular effective temperatures and bolometric luminosities - are excellent samples for testing stellar evolution models with a variety of metallicities and physical parameters (including current treatments of mass loss, rotation, and the effects of binary evolution, e.g. Ekstr\"om et al.\ 2012, Eldridge et al.\ 2017). Well-characterized RSG samples are also valuable when considering the role of these stars as core-collapse progenitors. Deep ground-based and {\it Hubble} Space Telescope photometry has been used to identify RSGs as the direct progenitors of Type II-P supernovae (e.g. Mattila et al.\ 2010, Van Dyk et al.\ 2012a, b; Maund et al.\ 2014; Smartt 2015) and the potential progenitors of black holes formed via direct collapse (Adams et al.\ 2017). Smartt et al.\ (2009) and Eldridge et al.\ (2013) estimated that such RSG progenitor detections are feasible out to $\sim$28 Mpc with {\it Hubble}. Using these pre-core-collapse detections to estimate the stars' initial masses and evolutionary states is currently our most direct means of studying the terminal evolution of massive stars.

Identifying RSGs in imaging is largely a straightforward means of applying color and magnitude cuts to select the brightest and reddest stars in a photometric sample. In the optical this usually corresponds to optical colors of ($V-R$)$_0 \ge 0.6$ (e.g. Massey \& Olsen 2003, Levesque \& Massey 2012) and magnitude cuts corresponding to a luminosity of log($L/L_{\odot}$)$\sim4.0$ (though it is worth noting that this is a conservative limit, often adopted to avoid contamination from ``super"-AGB stars, e.g. Levesque 2017). In the near- and mid-IR constraints on color parameters such as $Q1=(J-H) - 1.8 \times (H-K_s)$ (Negueruela \& Schurch 2007) and $Q2=(J-K_s)-2.69\times(K_s-[8.0])$ (Messineo et al.\ 2012) have been established and combined with single-color cuts (e.g. $J-H > 0.65$ and $H-K < 0.6$, Bonanos et al.\ 2009, 2010, Britavskiy et al.\ 2015) and magnitude cuts to identify extragalactic populations. Contamination from the plethora of foreground K and M dwarfs can be a significant problem with methods based solely on color and magnitude criteria; however, if $BVR$ photometry is available, dwarfs and supergiants can be separated on a $V-R$ vs. $B-V$ color-color diagram by exploiting surface gravity effects that lead to increased line blanketing (and thus decreased fluxes) in the blue spectra of RSGs (see Massey 1998 for discussion). No similar criteria for removing foreground contaminants have been identified in the near- or mid-IR (though it is worth noting that foreground contamination in mid-IR-selected samples is minimal: RSGs are bright in the mid-IR because of significant mass loss and dust production, a trait not shared with foreground dwarfs, e.g. Britavskiy et al.\ 2015).

With photometrically-identified RSG samples in hand, the logical next step is to determine these stars' effective temperatures and luminosities and to place them on the H-R diagram, where their positions can be compared to the predictions of stellar evolutionary theory and used to estimate additional parameters such as initial mass. Currently, optical spectroscopy is the most reliable means of determining RSG physical properties, particularly effective temperature ($T_{\rm eff}$). Fitting the $T_{\rm eff}$-sensitive TiO absorption bands in the optical produces the best agreement with the $T_{\rm eff}$ values predicted by stellar evolutionary theory, including the metallicity-dependent evolution of the Hayashi limit (e.g. Levesque et al. 2005, 2006; Massey et al.\ 2009; Levesque \& Massey 2012). Tabanero et al.\ (2018) recently used the equivalent widths of weak atomic absorption features produced by neutral metals to estimate $T_{\rm eff}$ in Magellanic Cloud RSGs and G-type supergiants, but the precision of this technique was insufficient for reproducing metallicity-dependent $T_{\rm eff}$-variations and is also less effective for M-type stars (see also Dorda et al.\ 2016). Some work has also focused on determining spectral types based on spectroscopic features in the near-IR, such at the CN band at 1.1$\mu$m, the first and second overtone CO bands at 2.3$\mu$m and 1.6$\mu$m respectively, and a featured dubbed the ``H-hump" by Davies et al.\ (2013) at 1.6$\mu$m, produced by a local H$-$ opacity minimum (e.g. Lancon et al.\ 2007, Davies et al.\ 2007, 2013, Rayner et al.\ 2009).

However, $T_{\rm eff}$ diagnostics using photometric colors are also highly effective. Calibrations relating $(V-K$) colors and $T_{\rm eff}$ are now available across a broad range of metallicities and agree with $T_{\rm eff}$ determined from TiO absorption band fitting (though $(V-K$) generally produces slightly warmer values; for more discussion see Levesque 2017). ($V-R$) colors are also a good means of estimating $T_{\rm eff}$; the shorter wavelength baseline renders it slightly less sensitive to $T_{\rm eff}$ but also less sensitive to reddening effects (Levesque et al.\ 2006), a particular concern in RSGs due to circumstellar dust. It is, however, important to note that the use of $V$-band RSG photometry in both of these diagnostics also renders them more challenging; RSGs are known to be variable in $V$ by up to a magnitude (e.g. Josselin et al.\ 2000), so accurate estimates of $T_{\rm eff}$ based on color require simultaneous photometry in both bands.

Concerns regarding optical variability and $T_{\rm eff}$ also apply when calculating RSG bolometric magnitudes ($M_{\rm bol}$). While RSGs are significantly variable in $V$, their $K$ magnitudes are much more stable (e.g. Josselin et al.\ 2000, Massey et al.\ 2009) and also much less sensitive to reddening ($A_K=0.11A_V$, Schlegel et al.\ 1998), making it the preferred band to use for estimating $M_{\rm bol}$. RSGs also have large and $T_{\rm eff}$-sensitive bolometric corrections in both $V$ and $K$, a consequence of spectral energy distributions that peak at $\sim$0.67-0.85$\mu$m and suffer substantial $T_{\rm eff}$-dependent line blanketing effects in the optical due to molecular absorption (e.g. Massey \& Olsen 2003, Levesque et al.\ 2005, 2006; see also Davies \& Beasor 2018).

Currently scheduled for launch in 2021, {\it JWST} is going to serve as a new and incredibly powerful tool for studying RSGs. The high luminosities and red spectral energy distribution (SED) peaks of these stars make them ideal targets for the telescope's IR focus, and the depth and resolution of {\it JWST}'s imaging instruments will make it possible to photometrically identify RSGs throughout the Local Volume (e.g. Jones et al.\ 2017). As a result, this data can be used to build up a tremendous database of pre-explosion imaging for future core-collapse transients (which themselves will be discovered at an unprecedented rate in the era of the Large Synoptic Survey Telescope) and an immense new observational sample for testing stellar evolution models. Making effective use of these new capabilities requires quantifiable photometric and spectroscopic diagnostics tailored for {\it JWST}. While methods for identifying candidate RSG populations will remain fairly straightforward (identifying the brightest and red stars in an imaging survey) the ability to interpret these observations - specifically, removing foreground contaminants and placing these stars on the H-R diagram - is a crucial step if we wish to make good and immediate use of {\it JWST} data.

This work simulates {\it JWST} near-IR photometry based on both models and observations of RSGs and presents the best color diagnostics for determining $T_{\rm eff}$, bolometric corrections tailored for the {\it JWST} near-IR filters, and color-color diagnostics that can be used to separate foreground dwarfs from background RSG samples. The models and data used to develop and test these diagnostics are presented in \S2, and the process of simulating {\it JWST} photometry is detailed in \S3. \S4 compares the two current sets of model atmospheres that are available for RSGs and identifies the models that are best suited for simulating near-IR photometry. \S5 presents color diagnostics for $T_{\rm eff}$ and bolometric corrections for determining $M_{\rm bol}$ and tests the efficacy of these methods at determine RSG physical properties is compared to current optical and near-IR techniques on the H-R diagram. \S5 also illustrates potential color-color methods for separating foreground and backgrounds samples in {\it JWST} near-IR observations. Finally, \S6 discusses these results and described potential future work on developing tools for studying RSGs with {\it JWST}.

\section{Data}
\subsection{Model Atmospheres}
\begin{figure*}
\center
\includegraphics[width=13cm]{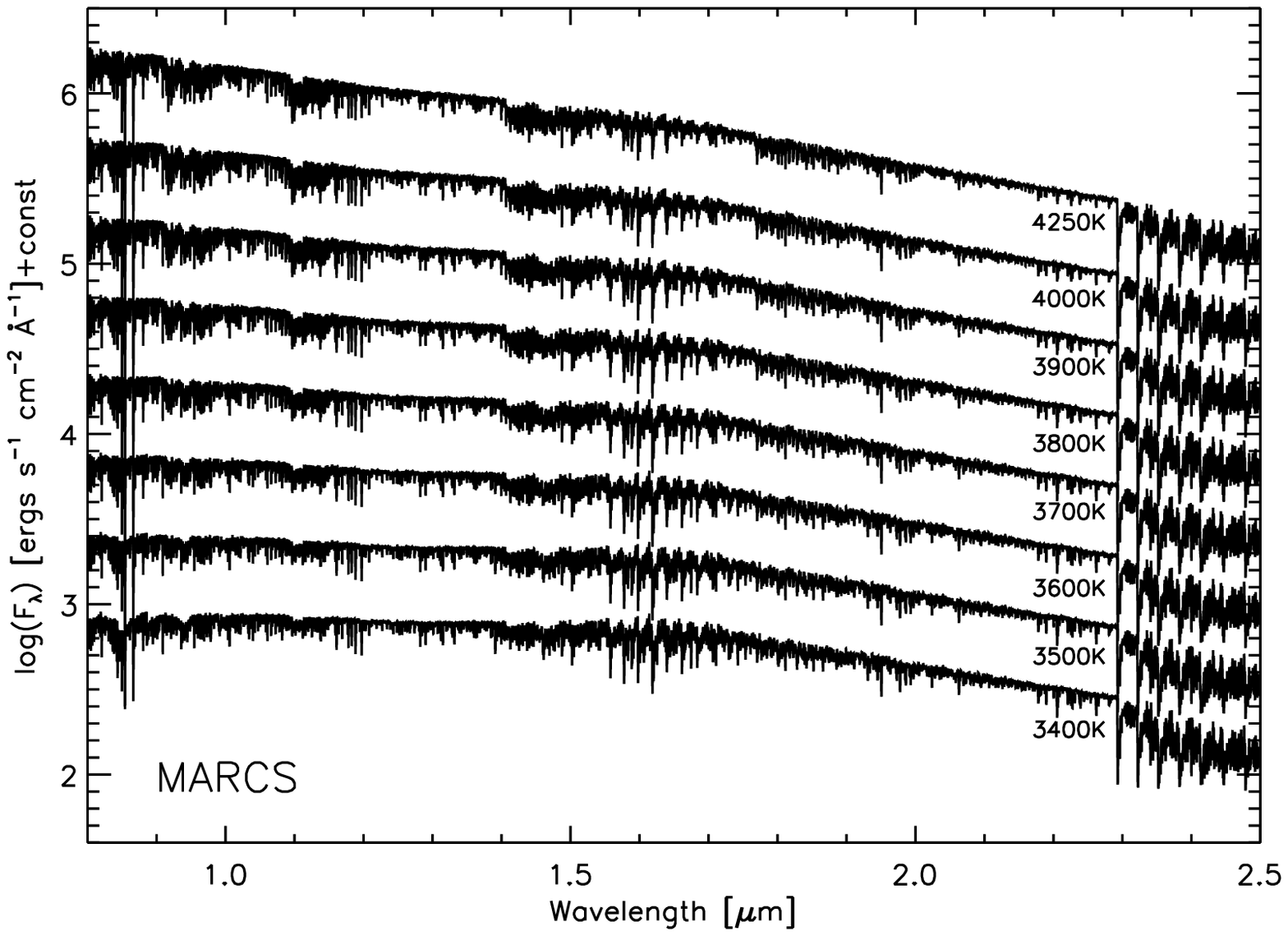}
\includegraphics[width=13cm]{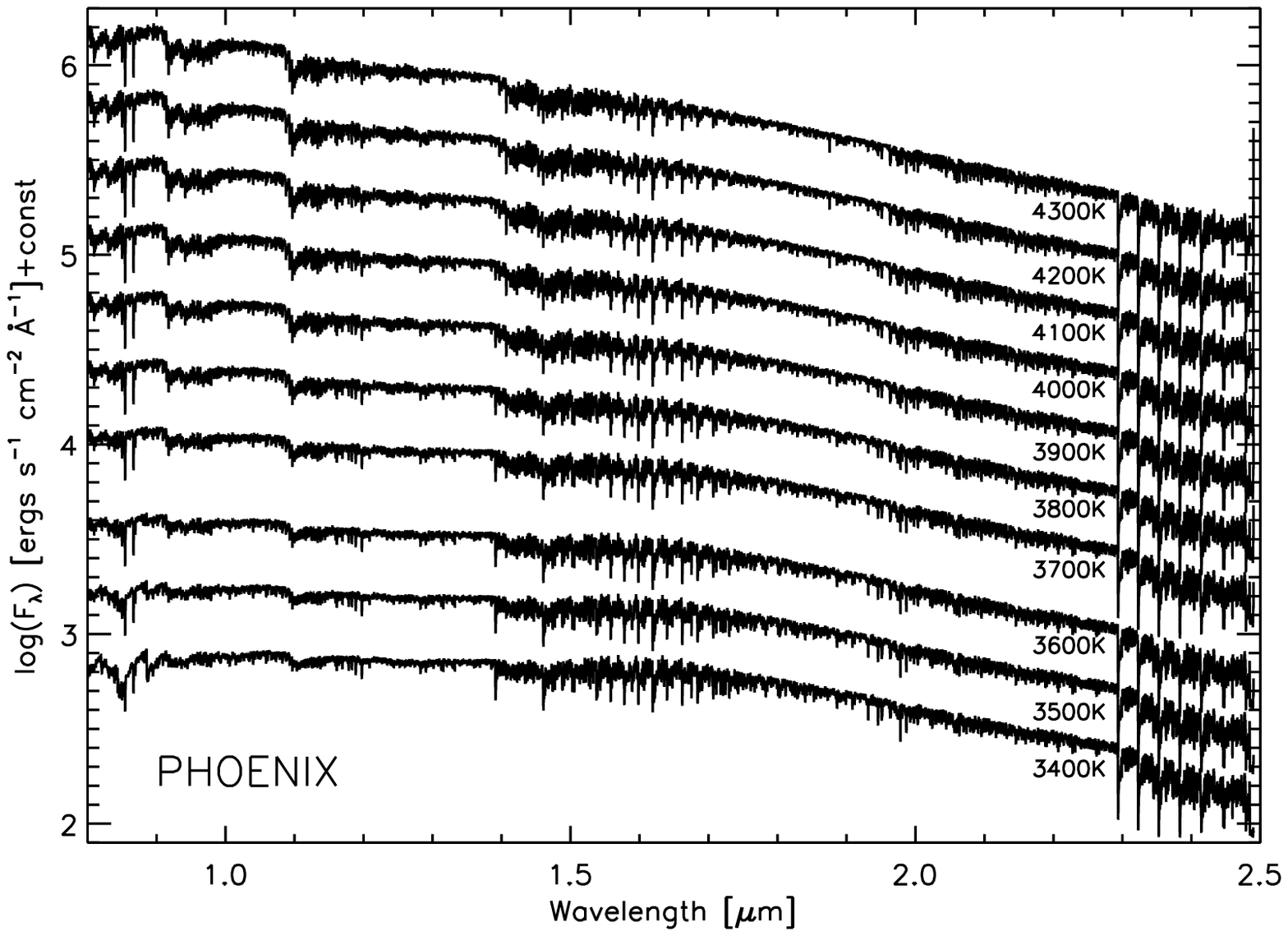} 
\caption{Synthetic RSG near-IR spectra produced by the MARCS (top) and PHOENIX (bottom) stellar atmosphere models, spanning the full range of RSG $T_{\rm eff}$ observed in the Milky Way (L05). For illustrative purposes the spectral resolutions of the two models have been matched at 1.5$\mu$m to $R\sim2700$, approximating the resolution of the {\it JWST}/NIRSpec high-resolution gratings.}
\end{figure*}

The envelopes of RSGs are extended and very cool, with high opacities due to the partial ionization of H$^-$ leading to large surface convection cells and sonic or even supersonic convective speeds. This makes their outer layers and observed spectra extremely difficult to model. While some progress has been made in producing three-dimensional models of RSG atmospheres (e.g. Chiavassa et al.\ 2009, 2011), most large grids of RSG atmosphere models make computationally-necessary approximations, generating one-dimensional spherically symmetric models that use mixing length theory to describe the effects of convection. The MARCS (e.g. Gustafsson et al.\ 2008) and PHOENIX (e.g. Lancon et al.\ 2007, Husser et al.\ 2013, Arroyo-Torres et al.\ 2013) model grids offer the best theoretical tools currently available for simulating the atmospheres of RSGs. This work therefore considers both the MARCS and PHOENIX stellar atmosphere models, comparing these models to each other and to existing RSG observations (see \S4).

The MARCS stellar atmosphere models used here generate synthetic spectra for solar-metallicity 15M$_{\odot}$ RSGs adopting local thermodynamic equilibrium (LTE), a spherical atmospheric geometry, and continuum and molecular line opacities from the literature, most notably TiO  opacities from Plez (1998) and Plez et al.\ (2003); for a more detailed discussion see Gustafsson et al.\ (2008). This work uses MARCS models with T$_{\rm eff}$s of 3400, 3500, 3600, 3700, 3800, 3900, 4000, and 4250 K, along with a log$(g)=0.0$ (with $g$ in cm$^2$ s$^-1$) and a microturbulence of 5 km s$^{-1}$, downloaded from the MARCS homepage\footnote{\texttt{http://marcs.astro.uu.se/index.php}}. MARCS models are computed across a very broad wavelength range, covering $\sim$0.13-20$\mu$m, with a constant resolution of $R=20,000$ (i.e., $\Delta \lambda$ increases as a function of wavelength). The MARCS model spectra are plotted in Figure 1 (top).

The PHOENIX stellar atmosphere models used here (e.g. Lancon et al.\ 2007) were computed using {\texttt{PHOENIX}} v13.11.00B, a general non-LTE atmosphere modeling package. These models also adopt a spherical geometry and solar metallicity, and use opacities computed from a master line list, including TiO lines from Schwenke (1998); for further discussion see Ku\u{c}inskas et al.\ (2005). This work uses PHOENIX models with 3400 $<$ T$_{\rm eff}$ $<$ 4300 K in 100 K increments, along with a log$(g)=0.0$ and a constant statistical velocity field of $\xi =$ 2 km s$^{-1}$ (which is treated as a microturbulence in the model calculations). The models were downloaded from the data made available publicly by Lancon et al.\ (2007), and cover $\sim$0.51-2.49$\mu$m with a constant $\Delta \lambda = 0.25$\AA. The PHOENIX model spectra are plotted in Figure 1 (bottom).
 
\subsection{Observed Spectra}
\begin{figure*}
\epsscale{1}
\includegraphics[width=17cm]{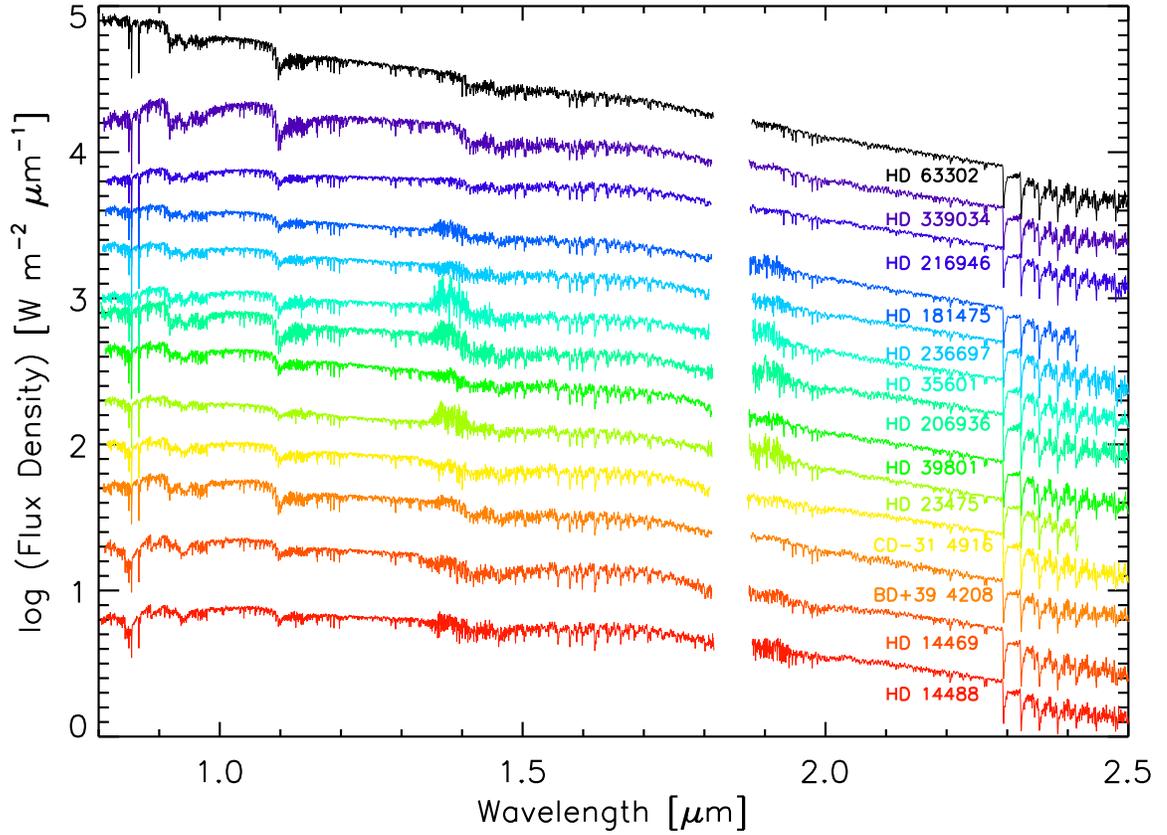}
\caption{Near-IR spectra from R09 of the 13 RSGs used as a comparison sample in this work (see Table 1). Spectra are plotted from top to bottom in order of decreasing $T_{\rm eff}$ as determined by L05. Note the coverage gap at $\sim$1.8-1.9$\mu$m.}
\end{figure*}
The infrared spectrophotometry used in this work is taken from the 3.0m NASA Infrared Telescope Facility (IRTF) spectral library for cool stars presented in Rayner et al.\ (2009; hereafter R09). The spectra were observed with the medium-resolution infrared spectrograph SpeX and span 0.8-5$\mu$m with $R \sim 2000$, with coverage gaps at $\sim$1.81-1.87$\mu$m, $\sim$2.54-2.83$\mu$m, 3.1-3.2$\mu$m, and 4.2-4.5$\mu$m. The continuum shape of the spectra was preserved during reduction, allowing them to be flux calibrated using accompanying photometry from the Two Micron All Sky Survey (2MASS; Skrutskie et al.\ 2006). For a complete discussion of the IRTF spectral library specifications and reduction procedures see R09.

The RSG sample has been restricted to stars that are included in both the R09 library {\it and} in the survey of Galactic RSGs presented in Levesque et al.\ (2005; hereafter L05). This made it possible to adopt measurements of the RSGs' physical properties based on well-established optical diagnostics (most notably T$_{\rm eff}$ based on the strength of the optical TiO absorption bands, as there is not yet an established spectroscopic T$_{\rm eff}$ diagnostic in the IR). A complete list of the RSGs used in this work and their physical properties from the literature is given in Table 1, and the RSGs' IR spectra are plotted in Figure 2.

\begin{deluxetable}{l l l c c c c c c c c}
\tabletypesize{\scriptsize}
\tablewidth{0pc}
\tablenum{1}
\tablecolumns{11}
\tablecaption{\label{tab:params} Red Supergiant Sample}
\tablehead{
\colhead{Star}
&\colhead{$\alpha_{2000}$}
&\colhead{$\delta_{2000}$}
&\multicolumn{2}{c}{Sp Type}
&\colhead{$(m-M)_0$\tablenotemark{a}}
&\colhead{$T_{\rm eff}$}
&\multicolumn{2}{c}{$A_V$}
&\colhead{$M_{\rm bol}$} 
&\colhead{Notes}\\ \cline{4-5} \cline{8-9}
\multicolumn{3}{c}{}
&\colhead{L05}
&\colhead{R09}
&\colhead{(mag)}
&\colhead{}
&\colhead{L05}
&\colhead{R09}
&\multicolumn{2}{c}{}
}
\startdata
HD 236697 &01 19 53.62 &+58 18 30.7 &M1.5 I &M0.5 I &11.9 &3700 &1.55 &1.5 &$-$6.25 &V466 Cas \\ 
HD 14469 &02 22 06.89 &+56 36 14.9 &M3-4 I &M3-4 I &11.4 &3575 &2.01 &1.5 &$-$7.64 &SU Per \\ 
HD 14488 &02 22 24.30 &+57 06 34.3 &M4 I &M3.5 I &11.4 &3550 &2.63 &1.6 &$-$8.15 &RS Per \\
HD 23475 &03 49 31.28 &+65 31 33.5 &M2.5 II &M2 II &6.90 &3625 &1.08 &0.9 &$-$5.02 &BE Cam \\ 
HD 35601 &05 27 10.22 &+29 55 15.8 &M1.5 I &M1.5 I &10.7 &3700 &2.01 &1.7 &$-$6.81 &\nodata \\ 
HD 39801 &05 55 10.31 &+07 24 25.4 &M2 I &M1-2 I &6.73 &3650 &0.62 &0.6 &$-$8.34&$\alpha$ Ori \\ 
CD $-$31$^{\circ}$ 4916 &07 41 02.63 &$-$31 40 59.1 &M2.5 I &M3 I &13.0 &3600 &2.01 &1.7 &$-$7.85/$-$6.69 &\nodata \\ 
HD 63302 &07 47 38.53 &$-$15 59 26.5 &K2 I &K1 I &9.22 &4100 &0.62 &1.7 &$-$4.46 &\nodata \\ 
HD 181475 &19 20 48.31 &$-$04 30 09.0 &K7 II &M1 II &9.15 &3700 &1.39 &1.6 &$-$5.03&\nodata \\ 
HD 339034 &19 50 11.93 &+24 55 24.2 &K3 I &M1 I &11.8 &4000 &5.27 &4.2 &$-$8.63 & Case 15 \\ 
BD +39$^{\circ}$ 4208 &20 28 50.59 &+39 58 54.4 &M3-4 I &M3 I &10.6 &3600 &4.49 &3.6 &$-$8.15 &RW Cyg \\ 
HD 206936 &21 43 30.46 &+58 46 48.1 &M1 I &M2 I &9.7 &3700 &2.01 &2.0 &$-$9.08 & $\mu$ Cep \\
HD 216946 &22 56 26.00 &+49 44 00.8 &M0 I &K5 I &8.9 &3800 &0.31 &0.5 &$-$5.50 &\nodata
\enddata	      	     
\tablenotetext{a}{Distance moduli are adopted from L05 with the following exceptions: HD 39801 (Betelgeuse; distance from Harper et al.\ 2017), and HD 23475, HD 63302, and HD 181475 (distance from the Hipparcos catalog; Van Leeuwen 2007). For the latter stars, distances determined from Hipparcos are consistent with their positions in {\it Gaia} DR2 when accounting for errors.}
\end{deluxetable}

The RSG sample here is representative of the range of $T_{\rm eff}$ and spectral types observed in the larger Milky Way RSG population. Galactic RSGs range from $\sim$4300 K (corresponding to a $\sim$K1-2 I spectral type) to $\sim$3400 K (corresponding to a $\sim$M5-6 I) spectral type but no cooler, a consequence of the minimum $T_{\rm eff}$ imposed by the Hayashi limit (though this limit shifts to even warmer $T_{\rm eff}$ at lower metallicities; see also \S6). This agrees with the predictions of stellar evolutionary theory (e.g. Ekstr\"om et al.\ 2012) and is reflected in the spectral types assigned by surveys of solar-metallicity RSGs using a variety of spectral type diagnostics (e.g. Levesque et al.\ 2005, Figer et al.\ 2006, Davies et al.\ 2007, Schuster et al.\ 2009, Massey et al.\ 2009, Massey \& Evans 2016).

The spectra used in L05 and R09 have both been corrected for total line-of-sight reddening effects, including contributions from foreground, local (OB association), and circumstellar dust. In L05 $E(B-V)$ was one of the parameters determined from optical stellar atmosphere model fitting, and the data were corrected based on the Cardelli et al.\ (1989) reddening law. In R09 an $E(B-V)$ was adopted based on the expected $(B-V)_0$ for the star's spectral type, and the spectrum was corrected using the Fitzpatrick \& Massa (2007) reddening law.

For several stars there are disagreements between the spectral types determined by L05 and those adopted from the literature (Keenan \& McNeil 1989, Garcia 1989, Keenan \& Newsom 2000) by R09. Traditionally, RSG spectral types have been based on the depths of the TiO absorption bands between $\sim$0.47-0.71$\mu$m for M-type and late-K-type stars, and on the shape of the continuum, the strength of the CH G band at 0.43$\mu$m, and neutral metal line ratios for early-K stars (Jaschek \& Jaschek 1990, L05). Minor differences in spectral typing methods can account for small variations ($\sim$1 subtype); similar disparities were found in L05 when comparing spectral types from previous work. While a small subset of RSGs do display substantial variability in their spectral types (and, as a result, their apparent $T_{\rm eff}$) on timescales as short as months, the stars in this sample do not display the other traits associated with this variability such as atypically late spectral types and changes in their amount of circumstellar dust (see Levesque et al.\ 2007); indeed, the sample includes several stars that were identified as spectral standards by Morgan \& Keenan (1973). However, one star in the sample, HD 339034, could show some signs of genuine variability. L05 determined a spectral type of K3 I for this star, along with a T$_{\rm eff}$ of 4000 K (consistent with a mid-K spectral type), and found that it showed signs of substantial circumstellar dust. By contrast, the star was assigned a spectral type of M1 I in R09 based on Garcia (1989) (as noted in L05, Humphreys 1978 also identified this star as an M1 I), so it is possible that these variations in spectral type represent real spectroscopic changes in the star. Another star in this sample, CD $-31^{\circ}$ 4916, was assigned two $M_{\rm bol}$ values by L05 because of larger-than-typical discrepancies between M$_{\rm bol}$ as determined from the $V$ magnitude and a lower M$_{\rm bol}$ as determined from $K$ (see \S5.2); both values are given in Table 1, but for consistency with the rest of the sample the L05 $M_{\rm bol}$ based on $V$ is adopted for the rest of this work.
 
 \section{Simulating JWST/NIRCam Photometry}
 \begin{figure*}
\epsscale{1}
\includegraphics[width=17cm]{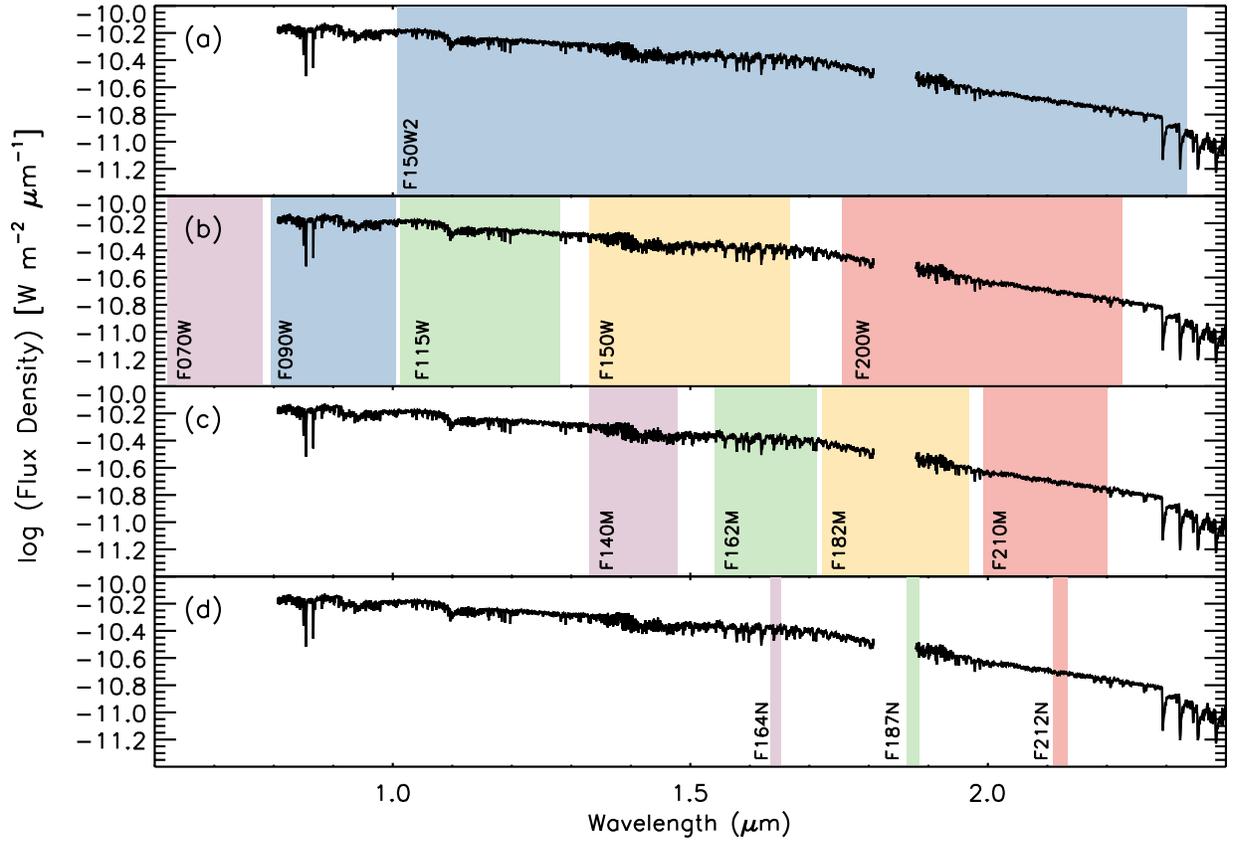}
\caption{An example R09 near-IR RSG spectrum (HD 236697) overlaid with the wavelength coverage of the {\it JWST}/NIRCam extra-wide (a), wide (b), medium (c), and narrow (d) filters in the near-IR.}
\end{figure*}

This work is focused on simulating future observations of RSGs using NIRCam on {\it JWST}. NIRCam is an infrared imager (Horner \& Rieke 2004) equipped with 13 bandpass filters in the short wavelength channel (near-IR; 0.6-2.3$\mu$m), including the ``extra-wide" F150W2 filter, five ``wide" filters (F070W, F090W, F115W, F150W, and F200W), four ``medium" filters (F140M, F162M, F182M, and F210M), and three ``narrow" filters (F164N, F187N, and F212N). The wavelength coverage for these filters is plotted in Figure 3 and compared to a sample R09 RSG spectrum.

The choice to concentrate this work on NIRCam's near-IR capabilities rather than its long wavelength channel (mid-IR) filters is in part a consequence of the significant effect that dust can have on RSG luminosities in the mid-IR. Circumstellar dust is quite common in RSGs (e.g. Massey et al.\ 2005, Verhoelst et al.\ 2009), and is most evident in the stars' high mid-IR luminosities. Observations in this regime can highlight cool circumstellar material produced by significant mass loss from RSGs (e.g. Britavskiy et al.\ 2014, Shenoy et al.\ 2016); however, stellar atmosphere models for RSGs do not currently include the effects of dust, and the effects of RSG circumstellar dust are difficult to quantify in a uniform manner due to uncertainties in the dust production mechanism, composition, and grain size (e.g. Speck et al.\ 2000, Massey et al.\ 2005, Levesque 2017). By comparison, in the near-IR the extinction effects of circumstellar dust are both well-understood (with the exception of extreme cases such as the dust-enshrouded OH/IR stars, e.g. Schuster et al.\ 2006; see also \S6) and much smaller, with $A_K=0.11A_V$ (Schlegel et al.\ 1998). The stars in our sample have total line-of-sight $A_V$ values that range from $0.31 < A_V < 5.27$, with a mean $A_V = 2.0$ (adopting values from L05 optical model fitting), which correspond to $0.03 < A_K < 0.58$ (mean $A_K = 0.23$).

Filter transmission throughput curves for the NIRCam short wavelength channel filters are available from the {\it JWST} website\footnote{\texttt{https://jwst-docs.stsci.edu/display/JTI/NIRCam+Filters}}. The throughput curves used here are the \texttt{modAB\_mean/nrc\_plus\_ote} files, which are the averaged throughput curves for NIRCam's module A and B (the two modules that make up the instrument, with nearly identical optics and detectors, to provide redundancy and double the available field of view). The modules' throughput curves were computed to include the effects of filter transmission, quantum efficiency, the dichroic beam splitter, and the effects of both the NIRCam and {\it JWST} optics. These throughput curves were combined with both the model and observed RSG SEDs to produce simulated NIRCam photometry. This was done for each SED and filter by stepping through the SED by wavelength, using the filter dispersion and throughput percentile at each step to determine the contribution of the SED to the total filter flux. The sum of the contribution at each step gives the total flux in that filter from the SED, which is then divided by the total area of the filter throughput curve to normalize by the passband. Taking the normalized filter flux $F_{\rm filter}$, $m_{\rm filter}=-2.5$log$(F_{\rm filter})-21.1$ then yields the magnitude in the filter following the STmag system (Stone 1996).

Photometry for a given filter was {\it not} determined in cases where the SED coverage of the relevant wavelength range was incomplete. As a result of the coverage gaps in the R09 spectra (see \S2.1), simulated photometry for these data was not generated for the F070W, F200W, F182M, and F187N filters. It is worth noting that the blue limit of the R09 data ($\sim$0.808$\mu$m on average) falls just within the wavelength range of the F090W filter (which covers 0.795-1.005$\mu$m); however, considering the small wavelength window and noting that the throughput in the F090W filter is lowest at the blue end, the impact on the resulting synthetic photometry is expected to be minimal.

\section{Comparison of RSG Stellar Atmosphere Models}
As a first step in generating useful {\it JWST/NIRCam} photometry diagnostics for RSGs, it is important to determine which models - MARCS or PHOENIX - are more effective at simulating observed RSG data in the near-IR. The MARCS models show generally excellent agreement with optical spectroscopy (Levesque et al.\ 2005), but matching both optical and IR regions of the SEDs with a single MARCS model is difficult; models that yield good agreement with the flux and TiO band depths in the optical tend to underestimate the strength of the near-IR spectrum (Davies et al.\ 2013). By contrast, the PHOENIX models show good spectroscopic agreement in the near-IR (Lancon et al.\ 2007), but their optical spectroscopic agreement is less satisfactory (for example, PHOENIX RSG spectra over-predict the presence and strength of a molecular absorption feature at $\sim$6500\AA\ that is not seen in observed RSG spectra).

For the purposes of this work, it is sufficient to determine which of the two models shows better agreement with the R09 data by comparing their simulated {\it JWST}/NIRCam photometry. Figure 4 plots $T_{\rm eff}$ against color for the data, MARCS, and PHOENIX models for every color index combination available (given the wavelength limitations of the R09 spectra). As expected, the R09 data points show a much broader scatter than those generated by the models; however, linear best fits to the R09 data show excellent agreement with the overall predictions of the models regarding the color indices' evolution with $T_{\rm eff}$ (for more discussion see \S5.1). The models show good agreement with each other and with the data for the simulated wide filter photometry; however, Figure 4 demonstrates that the PHOENIX models show generally better photometric agreement with the R09 data for the medium and narrow filter colors. In the F162M$-$F210M and F164N$-$F212N colors in particular the MARCS models are redder than the data, while the PHOENIX models show excellent agreement with the data's approximated linear correlation between temperature and color. These differences could arise from variations in how the models simulate RSG atmospheres and how they treat continuum and line opacities in the IR. In addition, the MARCS group explicitly notes that their model spectra give statistical estimates of the model surface fluxes at each wavelength step rather than full synthetic spectra; as a result, while these models are sufficient for generating synthetic {\it broadband} photometry, medium and narrow filter photometry based on the MARCS models may prove less effective (B. Plez, private communication). As a result, the model photometry used for the remainder of this work is based on the PHOENIX models.

\begin{figure*}
\center
\includegraphics[width=6.5cm]{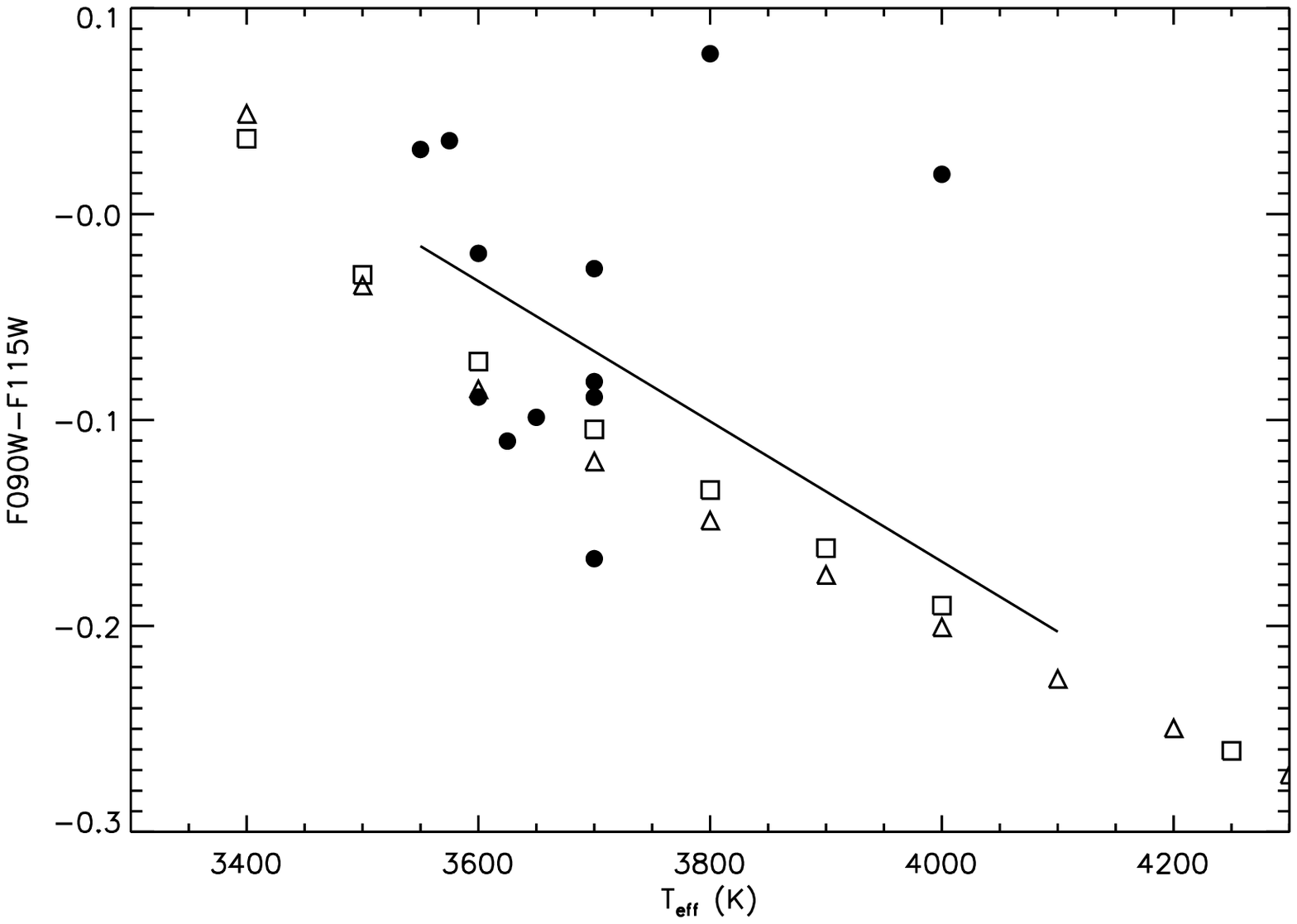}
\includegraphics[width=6.5cm]{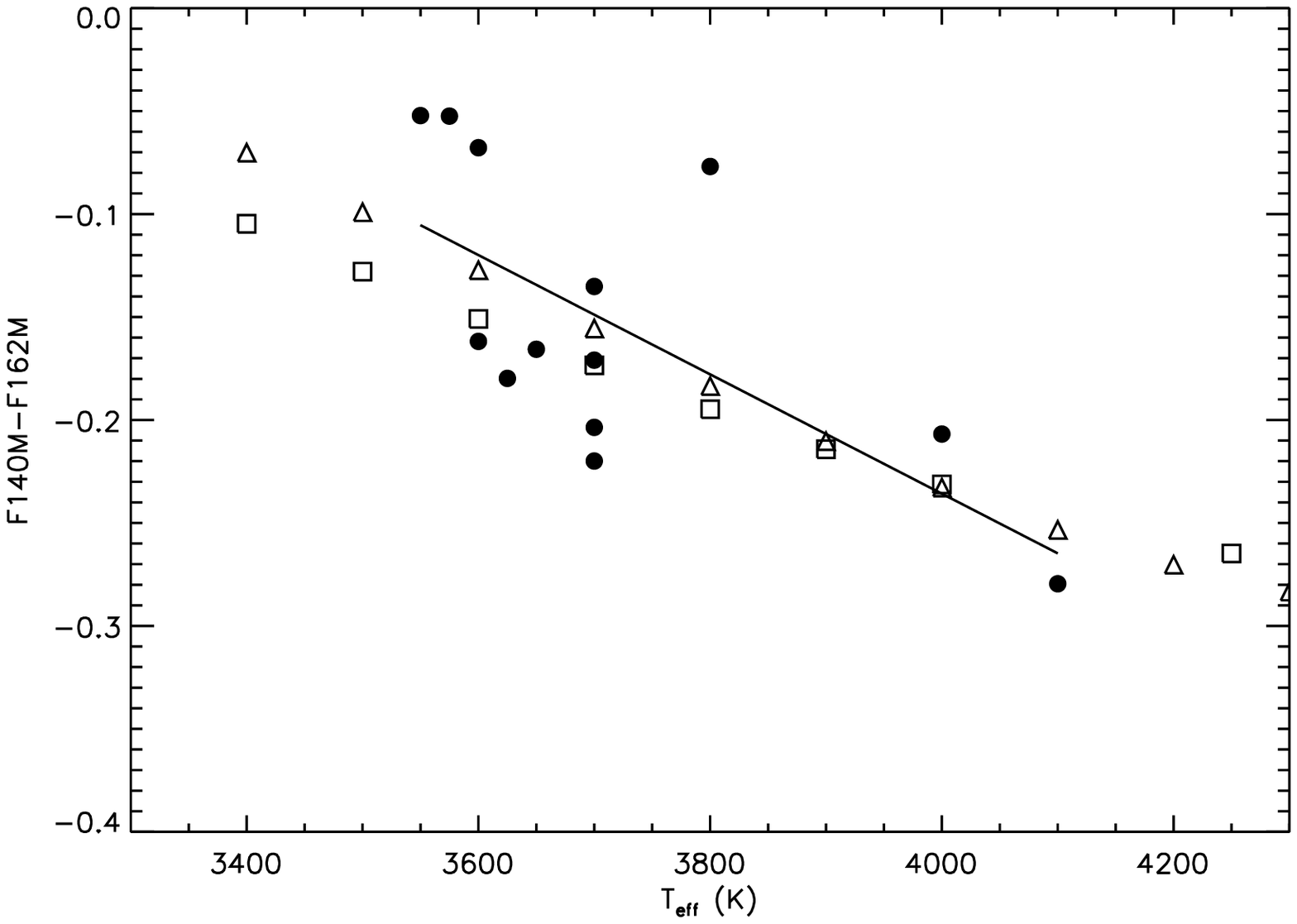}
\includegraphics[width=6.5cm]{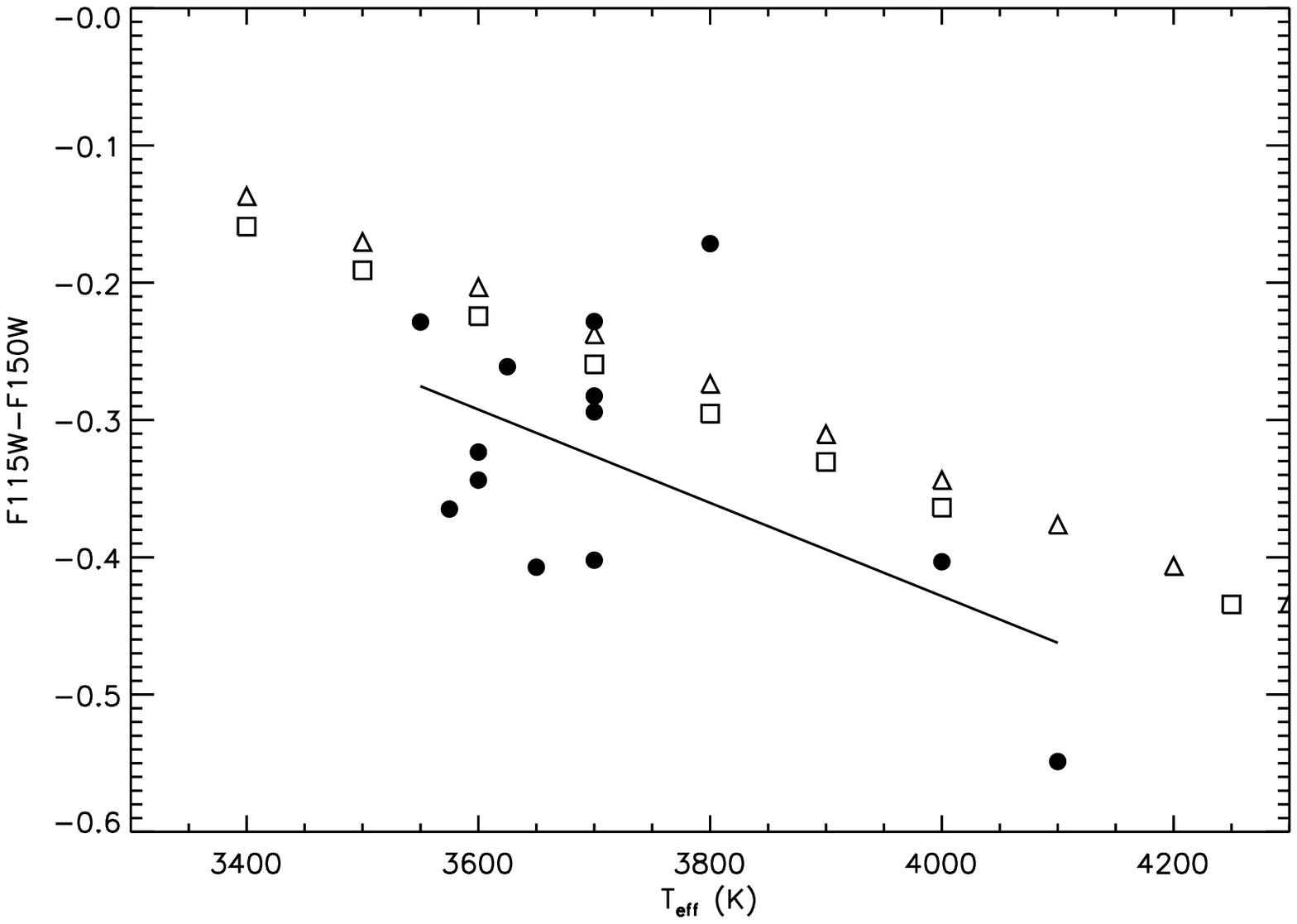}
\includegraphics[width=6.5cm]{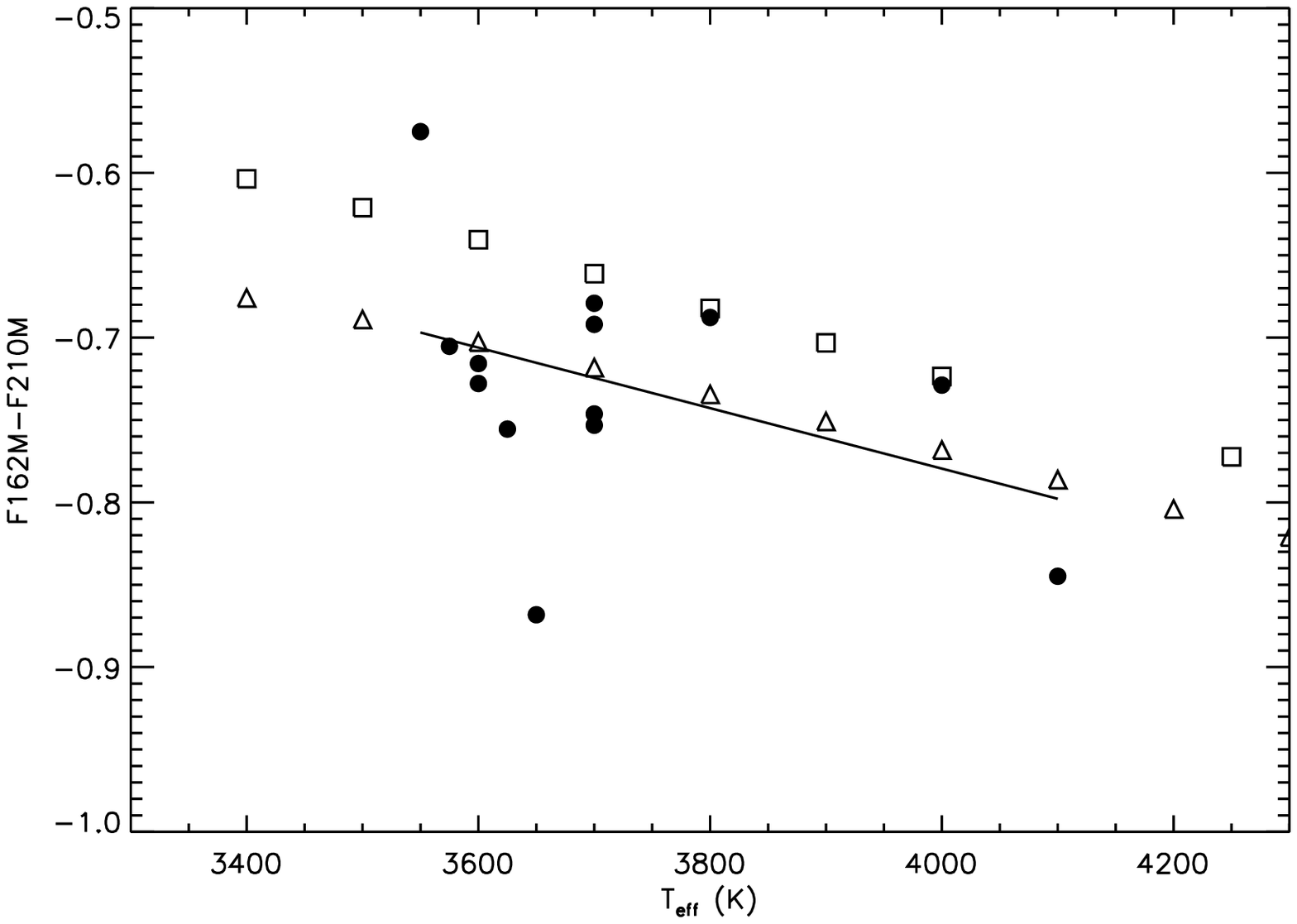}
\includegraphics[width=6.5cm]{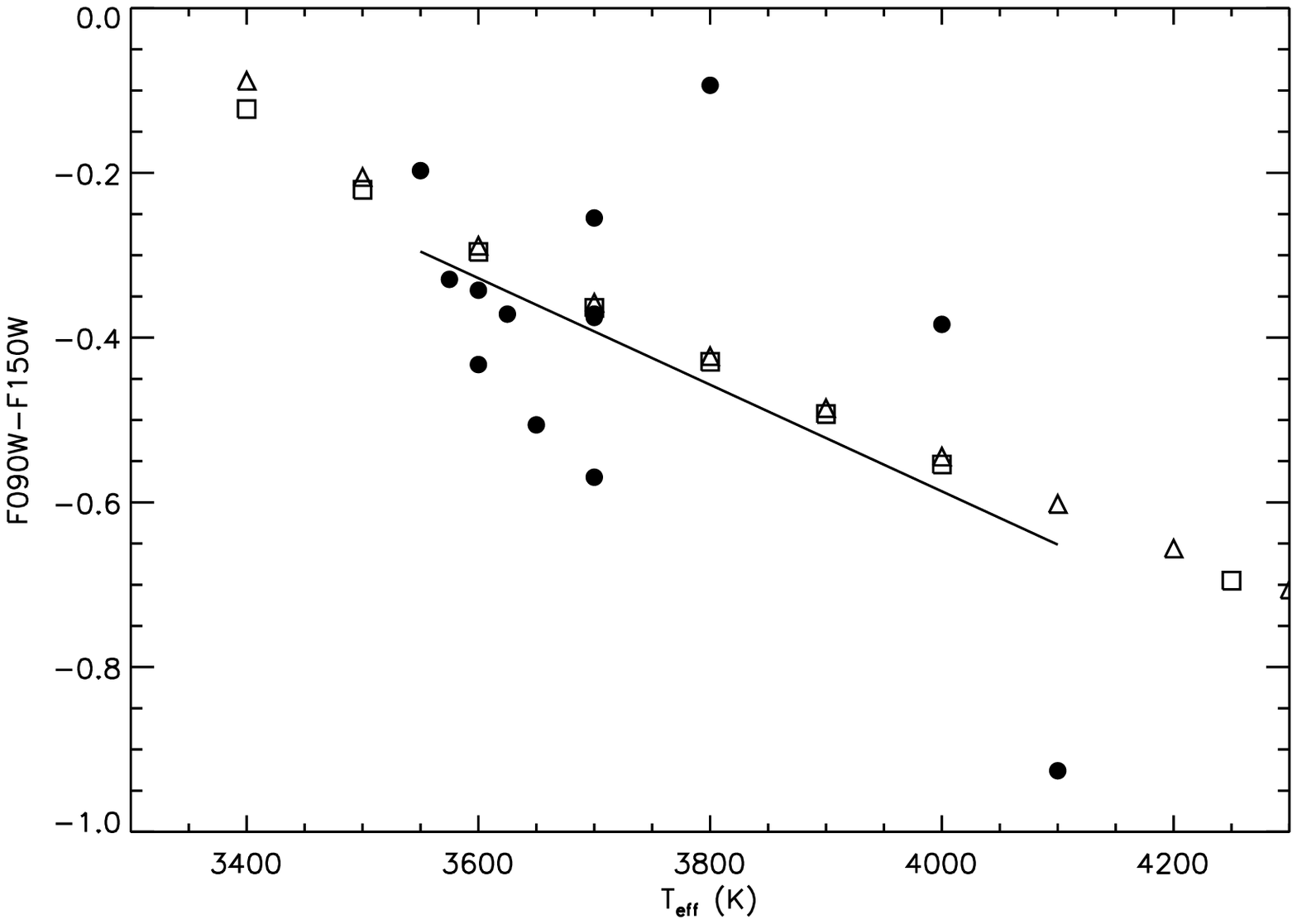}
\includegraphics[width=6.5cm]{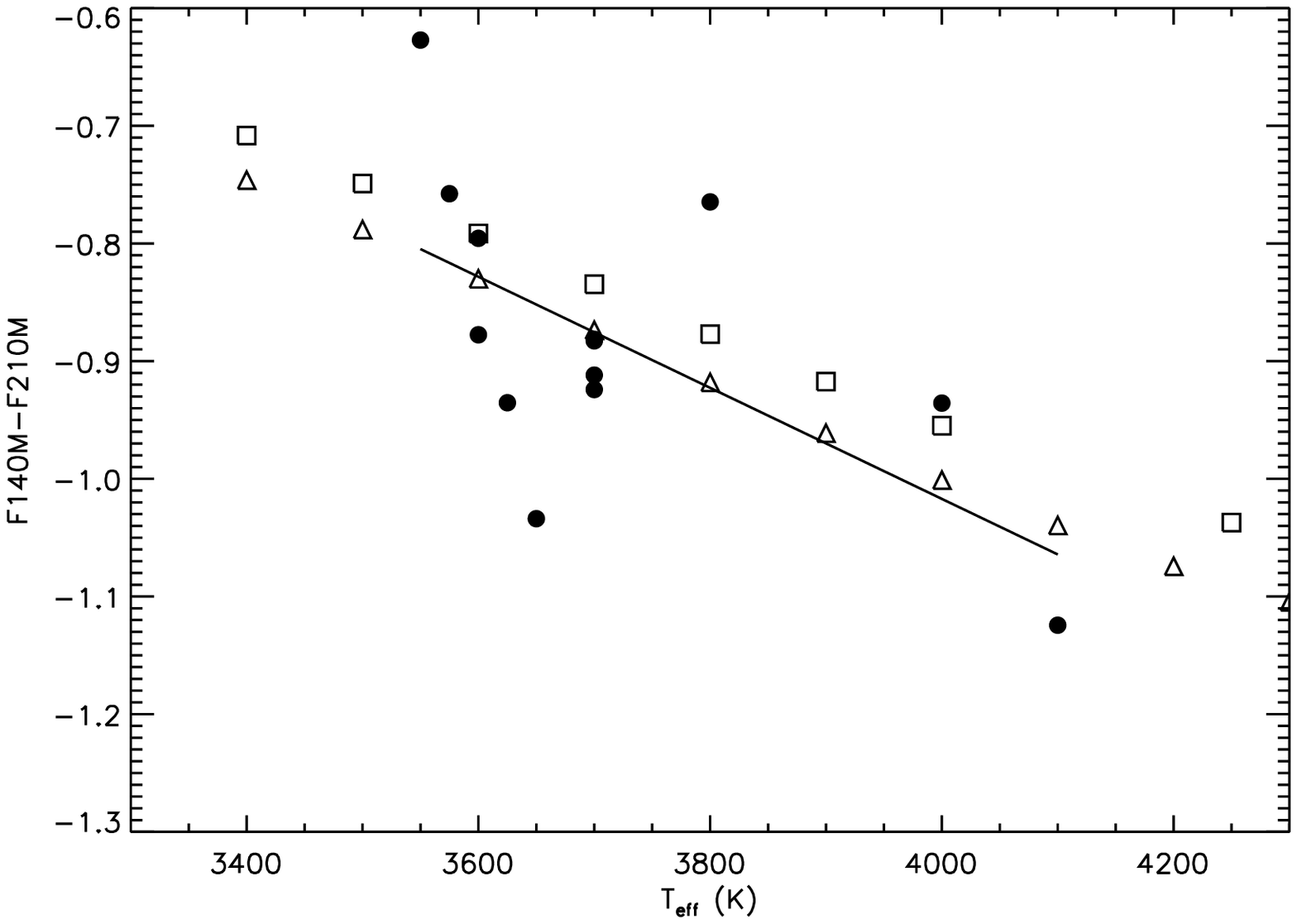}
\includegraphics[width=6.5cm]{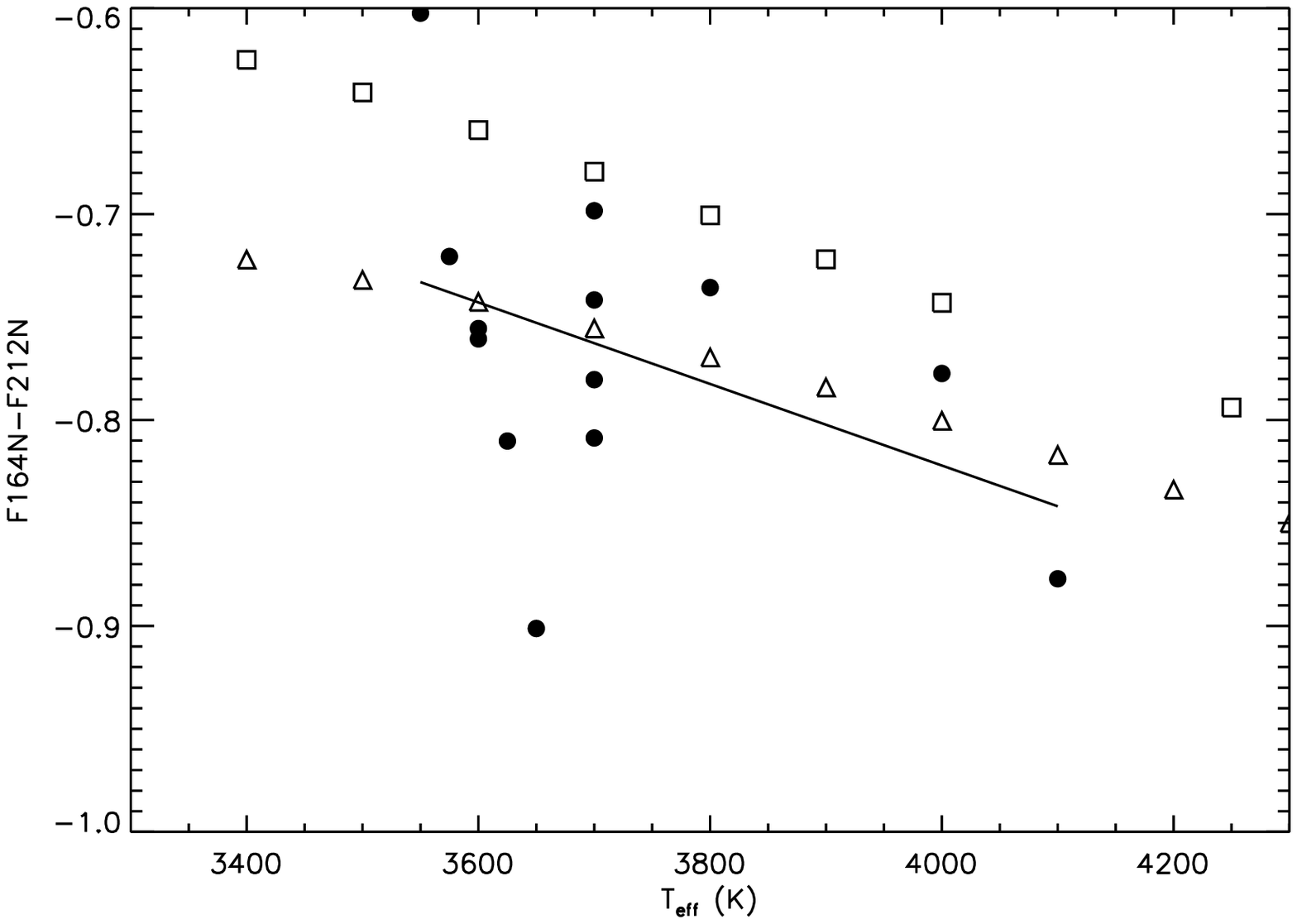}
\caption{$T_{\rm eff}$ vs. synthetic JWST/NIRCam near-IR photometry colors for the wide (left), medium (right), and narrow (bottom) near-IR filters covered by the R09 observations' wavelength range. The data points represent simulated photometry for the PHOENIX (triangles) and MARCS (squares) stellar atmosphere models along with the R09 supergiant spectra (filled circles). Linear best fits to the R09 data points are plotted as solid lines.}
\end{figure*}

It is worth noting that the $T_{\rm eff}$ adopted for the R09 data in this comparison are based on the L05 work, which compared the strengths of the TiO bands in the data to those predicted by the MARCS models to determine $T_{\rm eff}$. However, determining $T_{\rm eff}$ in the same manner (using the optical spectra and the TiO band strengths) using the PHOENIX model spectra produced very similar results, with the PHOENIX model fits yielding slightly warmer $T_{\rm eff}$ ($\sim$50K) for the R09 sample on average (for more discussion, see \S6). Regardless of which models are used to determine $T_{\rm eff}$ for the R09 RSGs, the PHOENIX models still show better agreement with the data.

\section{Photometric RSG Diagnostics for JWST}
\subsection{Effective Temperature}
In its simplest form, determining the $T_{\rm eff}$ of a RSG based on its photometric color is based on the star's blackbody spectrum, estimating the shape of the SED over the wavelength baseline bracketed by the chosen color index. However, in practice this is complicated by the substantial amount of line blanketing in these stars from molecular and atomic absorption (requiring that relations between $T_{\rm eff}$ and color index be based on actual SEDs rather than a simple blackbody curve). Currently, $(V-K)$ is the best color index for determining $T_{\rm eff}$ in RSGs, though $(V-R)$ is also effective despite a shorter wavelength baseline and corresponding weaker correlation with $T_{\rm eff}$ (Levesque et al.] 2006; see also \S1). It is also worth noting that these color indices are arguably superior to spectroscopic methods when determining $T_{\rm eff}$ in warmer RSGs, which have spectral types of early- to mid-K and thus fairly barren optical spectra that lack high-S/N $T_{\rm eff}$-sensitive spectral features (see Levesque \& Massey 2012, Levesque 2017).

This work examined the full set of color indices available from our synthetic {\it JWST} photometry to identify which colors are the most sensitive to $T_{\rm eff}$  in RSGs. In this context, ``most sensitive" corresponds to color indices that span the broadest range of magnitudes over the $T_{\rm eff}$ range of the model spectra (3400-4300 K for the PHOENIX models). The color-vs.-$T_{\rm eff}$ relation was computed for every pair of filters that could produce a traditional color index (``blue - red", where the bluer filter was established by comparing the two filters' mean wavelength coverage). This includes colors that combine extra-wide, wide, medium, and narrow filters.

The best color indices for determining $T_{\rm eff}$ in RSGs are plotted in Figure 5. As expected, the most $T_{\rm eff}$-sensitive color indices are those that span the broadest wavelength baseline (and are therefore the most sensitive to changes in the overall shape of the SED). The best color indices for determining $T_{\rm eff}$ also use the bluest filter, F070W, thus sampling a region of the spectrum that is particularly sensitive to $T_{\rm eff}$. The F070W filter covers the 0.621$\mu$m-0.781$\mu$m wavelength range; as a result, F070W magnitude will decrease with $T_{\rm eff}$ due to the rightward shift of the peak wavelength (which moves from $\sim$0.675$\mu$m at 4300 K to $\sim$0.853$\mu$m at 3400 K) as well as the increasing strength of the 0.616$\mu$m, 0.666$\mu$m, and 0.705$\mu$m TiO bands, which collectively decrease the stellar flux in this band's wavelength range. The F090W filter (spanning 0.795-1.005$\mu$m) is also an effective blue filter in $T_{\rm eff}$-sensitive color indices (offering a wavelength baseline that is almost as long and including the 0.843$\mu$m TiO band), though not to the same extent as F070W.

(F070W-F200W) is the best color index available in the {\it JWST}/NIRCam near-IR filters for determining $T_{\rm eff}$ in RSGs. The best-fit polynomial relation between (F070W-F200W) and $T_{\rm eff}$ can be expressed as:
\begin{equation}
T_{\rm eff} = 3470.34 - 244.617x + 58.3913x^2 -61.2692x^3 (\pm \sigma)
\end{equation}
where $x = (F070W-F200W)$ and $\sigma=0.748$. Best-fit polynomial coefficients for the full set of $T_{\rm eff}$-sensitive color indices plotted in Figure 5 are given in Table 2.

\begin{figure*}
\center
\includegraphics[width=9cm]{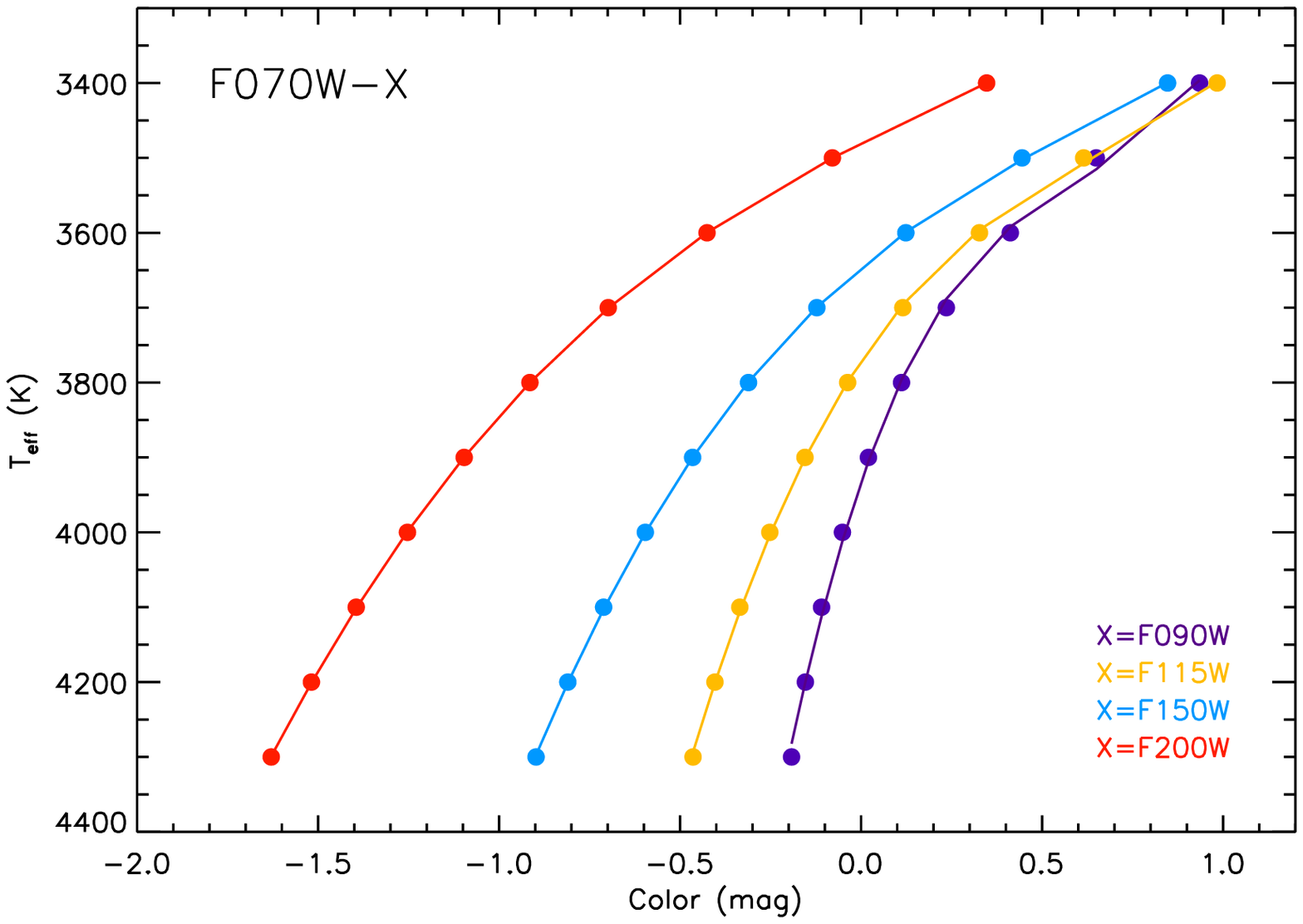}
\includegraphics[width=9cm]{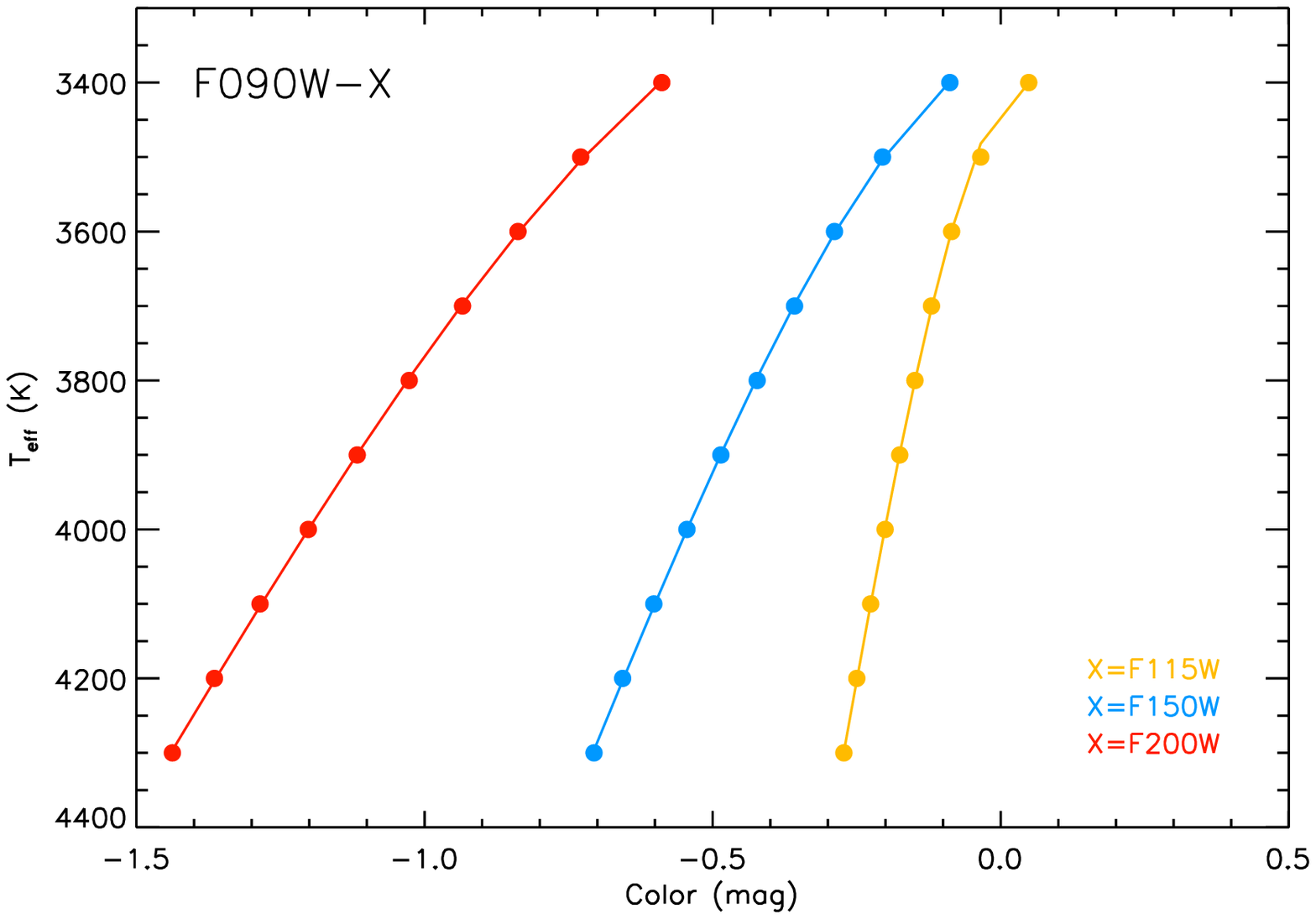}
\includegraphics[width=9cm]{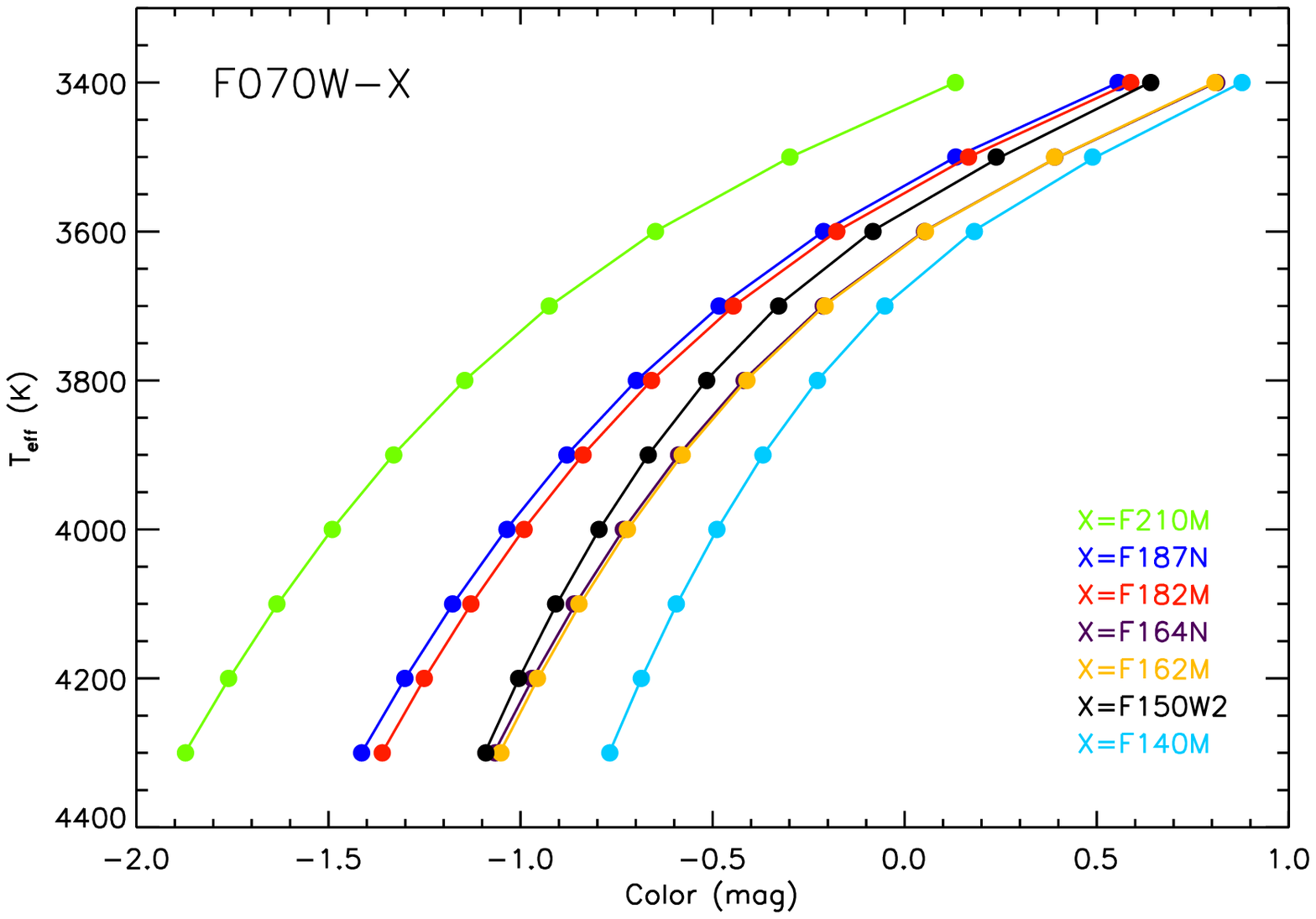}
\caption{A comparison of the strongest color-$T_{\rm eff}$ diagnostics in the NIRCam short wavelength channel filters, as computed from the PHOENIX models. The top panel shows the color vs. $T_{\rm eff}$ relations for wide-filter color indices using F070W as the blue filter, while the center panel shows the same for color indices using F090W as the blue filter. The bottom panel adopts F070W as the blue filter and illustrates color indices computed using the extra-wide, medium, and narrow NIRCam filters.}
\end{figure*}

\begin{deluxetable}{l c c c c c}
\tabletypesize{\scriptsize}
\tablewidth{0pc}
\tablenum{2}
\tablecolumns{6}
\tablecaption{\label{tab:params} Coefficients for $T_{\rm eff}$-sensitive {\it JWST}/NIRCam Color Indices\tablenotemark{a}} 
\tablehead{
\colhead{Color}
&\colhead{A}
&\colhead{B}
&\colhead{C}
&\colhead{D}
&\colhead{$\sigma$} \\
}
\startdata
F070W$-$F200W  &    3480.34  &   $-$244.617  &    58.3913  &   $-$61.2692   &  0.748001 \\
F070W$-$F150W  &    3645.72  &   $-$405.529   &   242.819  &   $-$126.867   &   1.69667 \\
F070W$-$F115W  &    3770.65  &   $-$725.220  &    700.140  &   $-$353.292   &   4.62617 \\
F070W$-$F090W  &    3933.39  &   $-$1422.28  &    1861.87  &   $-$1021.25   &   11.4175 \\
& & & & &  \\
F090W$-$F200W  &    3267.89  &    302.255  &    1016.07  &    214.246   &   3.05234 \\ 
F090W$-$F150W  &    3350.31  &   $-$381.207  &    1943.62  &    828.737   &   2.97820 \\
F090W$-$F115W  &    3429.27  &   $-$1155.49   &   10809.8   &   12193.8    &  1.85255 \\
& & & & &  \\
F070W$-$F210M  &    3429.54  &   $-$226.194  &    17.9169  &   $-$58.3301   &  0.951134 \\
F070W$-$F187N  &    3535.50  &   $-$280.337   &   99.1395   &  $-$60.0995    & 0.647344 \\
F070W$-$F182M  &    3545.15  &   $-$287.390  &    107.616  &   $-$65.3391   &  0.647259 \\
F070W$-$F164N  &    3616.79  &   $-$346.519  &    173.321  &   $-$93.9130   &   1.88444  \\
F070W$-$F162M  &    3617.78  &   $-$350.867  &    178.364  &   $-$97.4693   &   1.92625 \\
F070W$-$F150W2  &  3571.22  &   $-$319.076  &    167.441  &   $-$137.972   &   1.99693 \\
F070W$-$F140M  &    3674.35  &   $-$470.665  &    321.640  &   $-$161.760   &   1.38912
\enddata	      	    
\tablenotetext{a}{Equations take the form of $T_{\rm eff}$ = A + Bx + Cx$^2$ + Dx$^3$, where x is the value of the color index, with an error of $\sigma$.}  	
\end{deluxetable}

\subsection{Bolometric Magnitude}
Along with $T_{\rm eff}$, $M_{\rm bol}$ is the other key physical parameter needed to place RSGs on the H-R diagram. With the placement of these stars at or near the Hayashi limit, $M_{\rm bol}$ is also the key parameter dictating the assumed initial mass of RSGs when compared to the predictions of stellar evolutionary tracks. Bolometric corrections (BCs) are, in turn, a crucial ingredient in determining $M_{\rm bol}$ from photometry of RSGs in both the optical {\it and} near-IR. Blackbody SEDs for RSGs peak at $\sim$0.675$\mu$m-0.853$\mu$m, falling at the red end of the standard optical wavelength range and just to the blue of the near-IR range, so in both regimes the BCs will, for most bands, be significant and very $T_{\rm eff}$-sensitive. L05 and Levesque et al.\ (2006) both note the strong $T_{\rm eff}$ dependence of $V$-band BCs, with the value of $BC_V$ decreasing (becoming more negative) as $T_{\rm eff}$ decreases (i.e., the SED peak is moving further from the $V$ band so a more substantial $BC$ is required). Determining $M_{\rm bol}$ based on $V$ magnitude is further complicated by these stars' photometric variability in the optical.

The near-IR, where RSGs show significantly less variability (e.g. Josselin et al.\ 2000, Massey et al.\ 2009), is a better choice when determing $M_{\rm bol}$, but quantifying the $T_{\rm eff}$ dependence of the BCs is still key. Equations for determining $T_{\rm eff}$-dependent BCs for the $K$ band across a range of metallicities are given in L05, Levesque et al.\ (2006), Massey et al.\ (2009), and Levesque \& Massey (2012). In this band the value of $BC_K$ {\it increases} (becomes more positive) as $T_{\rm eff}$ decreases (i.e., the SED peak is moving {\it closer} to the $K$ band so a {\it less} substantial BC is required). Davies et al.\ (2013) also estimated mean $BC_K$ values for the RSG populations of the LMC and SMC; however, these are calculated based on $T_{\rm eff}$ determined from fitting the overall shape of RSG SEDs in the optical and near-IR with the one-dimensional MARCS models, an approach that yields only a narrow range of $T_{\rm eff}$ values and does not reproduce, for example, the metallicity-dependent evolution of the Hayashi limit (for more discussion see Levesque 2017). More recently, Davies \& Beasor (2018) determined empirical BCs in $V$, $R$, $I$, and $K$ for RSGs based on observations of Galactic RSGs in clusters (although this work neglected the effects of circumstellar dust). These showed a correlation between BC and spectral type in agreement with previous work, with the value of the BC increasing at later spectral types (i.e. cooler stars) for the $K$ band and decreasing for the $V$, $R$, and $I$ bands.

The PHOENIX models can be used here to calculate BCs as a function of $T_{\rm eff}$ for the {\it JWST}/NIRCam near-IR filters. The models assume a stellar mass of 15$M_{\odot}$ and log$(g) = 0.0$, which corresponds to a stellar radius of $\sim641R_{\odot}$ following $g=GM/r^2$. For each model $T_{\rm eff}$ the associated model luminosity can then simply be calculated via $L=4\pi R^2\sigma T_{\rm eff}^4$, and the $M_{\rm bol}$ for the model in the STmag system is then given by $M_{\rm bol} = 4.74-2.5{\rm log}(L/L_{\odot})-21.1$. Finally, this can be combined with the synthetic {\it JWST} photometry from the PHOENIX models to determine the BC in each filter for each model $T_{\rm eff}$. It is important to keep the inherent assumptions of these models in mind, noting in particular that the BCs will vary for RSGs with different initial masses (i.e., most RSGs will have M$_{\rm i}<$ 15M$_{\odot}$) and different surface gravities (which range from $-$0.5 to 0.5 for most RSGs).

Table 3 quantifies the full set of $T_{\rm eff}$-dependent bolometric corrections in the STmag system; these relations are also plotted in Figure 6. As seen in previous work, the evolution of the BCs with $T_{\rm eff}$ diverges with filter wavelength, based on whether the filters are bluer (e.g. F070W and F090W) or redder than the RSG SEDs' peak wavelengths. The bolometric corrections in these filters are typically between $-$0.9 and $-$3.0 mag, and are unsurprisingly more negative (corresponding to a larger correction in luminosity) for the filters covering wavelengths that are farthest from the stars' SED peak. As a result, the F090W filter has both the smallest mean BC ($-$1.05) and, along with F115W (mean BC $= -1.20$), the least variation with $T_{\rm eff}$ ($\Delta$BC $= 0.16$ mag for the $T_{\rm eff}$ range represented here). While F070W also has a relatively small mean BC ($-1.24$), its variation with $T_{\rm eff}$ is notably larger ($\Delta$BC $= 1.29$), likely as a result of the TiO absorption bands exacerbating the decrease in F070W magnitude with $T_{\rm eff}$.

\begin{figure*}
\center
\includegraphics[width=17cm]{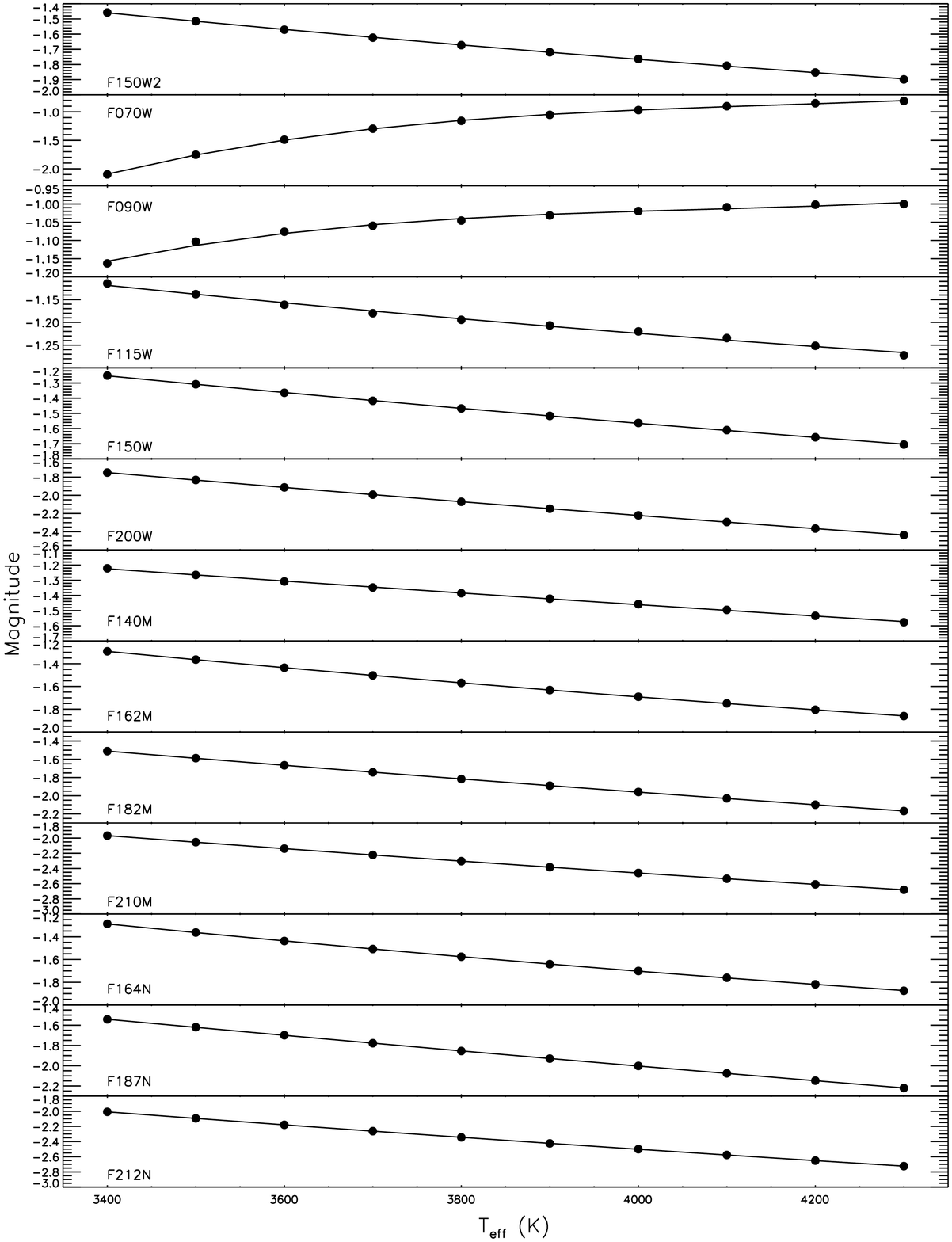}
\caption{Bolometric corrections computed using the PHOENIX stellar atmosphere models for the {\it JWST}/NIRCam near-IR filters.}
\end{figure*}

\begin{deluxetable}{l c c c c c}
\tabletypesize{\scriptsize}
\tablewidth{0pc}
\tablenum{3}
\tablecolumns{6}
\tablecaption{\label{tab:params} Coefficients for $T_{\rm eff}$-Dependent {\it JWST}/NIRCam Bolometric Corrections\tablenotemark{a}}
\tablehead{
\colhead{Filter}
&\colhead{A}
&\colhead{B}
&\colhead{C}
&\colhead{D}
&\colhead{$\sigma$}
}
\startdata
 F150W2 &        1.515 & $-$1.184  & 0.09089 &\nodata & 0.002  \\
 F070W &         $-$136.6 &   96.68 & $-$23.08 & 1.847  & 0.007 \\
 F090W &        $-$22.23 &   15.46 & $-$3.774 & 0.3083 &0.006 \\
 F115W &        0.009052 &$-$0.4642 & 0.03897 &\nodata &0.004 \\
 F150W &        1.483 &$-$1.045 & 0.07077 &\nodata &0.002 \\
 F200W &        1.962 & $-$1.353 & 0.0767 &\nodata & 0.001\\
 F140M &         0.4489 &$-$0.5752 & 0.02444 &\nodata &0.003 \\
 F162M &       2.650 & $-$1.578 & 0.1231&\nodata & 0.001\\
 F182M &      1.963 &  $-$1.252 & 0.06778 &\nodata & 0.001 \\
 F210M &        2.083 & $-$1.506 & 0.09251 &\nodata &0.001 \\
 F164N &       2.942 & $-$1.713 & 0.1379 &\nodata &0.001 \\
 F187N &        1.817 & $-$1.172 & 0.05428 &\nodata &0.001 \\
 F212N &        2.021 & $-$1.493 & 0.09061 &\nodata &0.001 
\enddata	      	    
\tablenotetext{a}{Equations take the form of $BC_{\rm filter}$ = A + Bx + Cx$^2$ + Dx$^3$, where x is $T_{\rm eff}$/1000, with an error of $\sigma$.}
\end{deluxetable}

\subsection{JWST Diagnostics on the H-R Diagram}
\begin{figure*}
\center
\includegraphics[width=10cm]{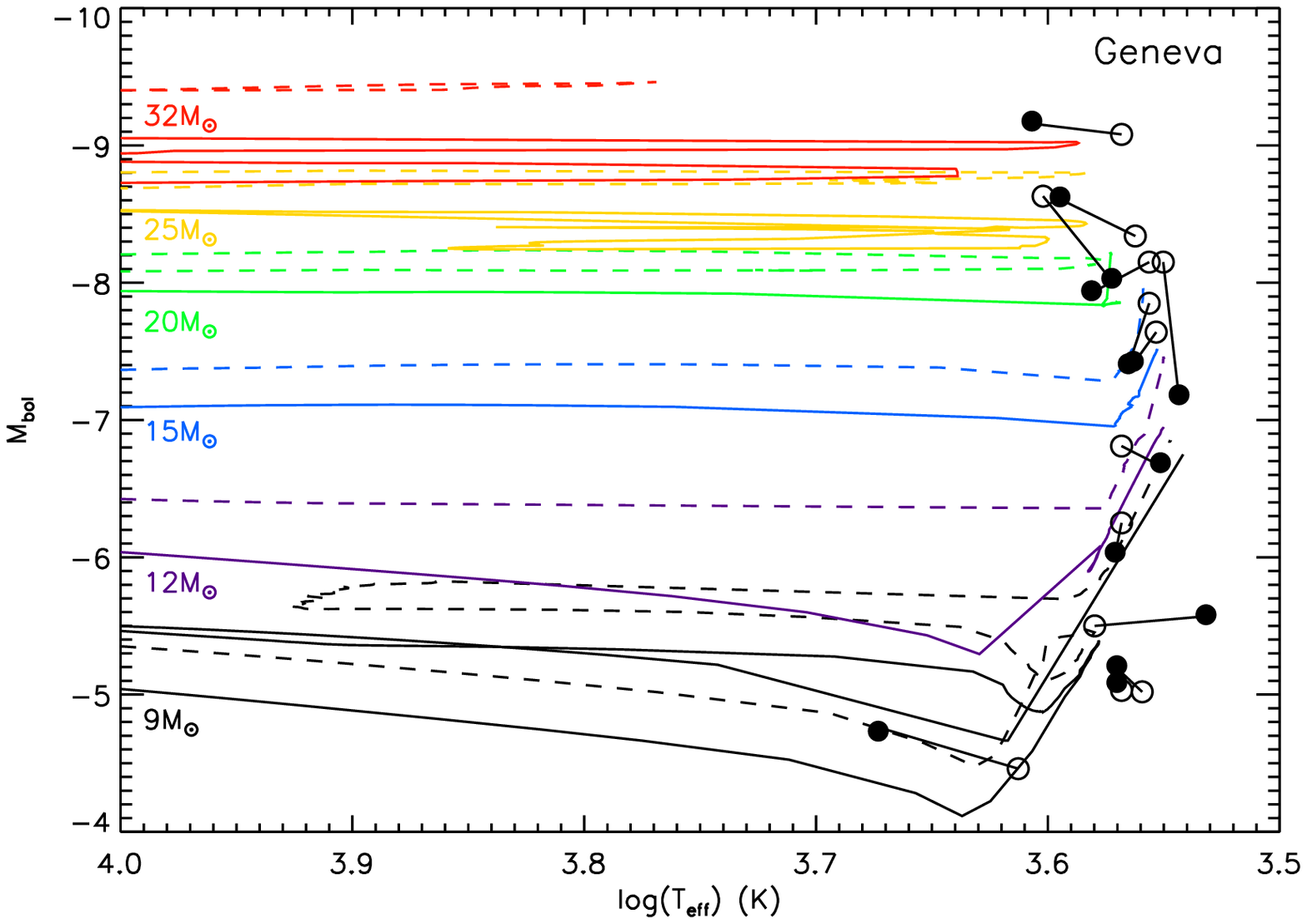}
\includegraphics[width=10cm]{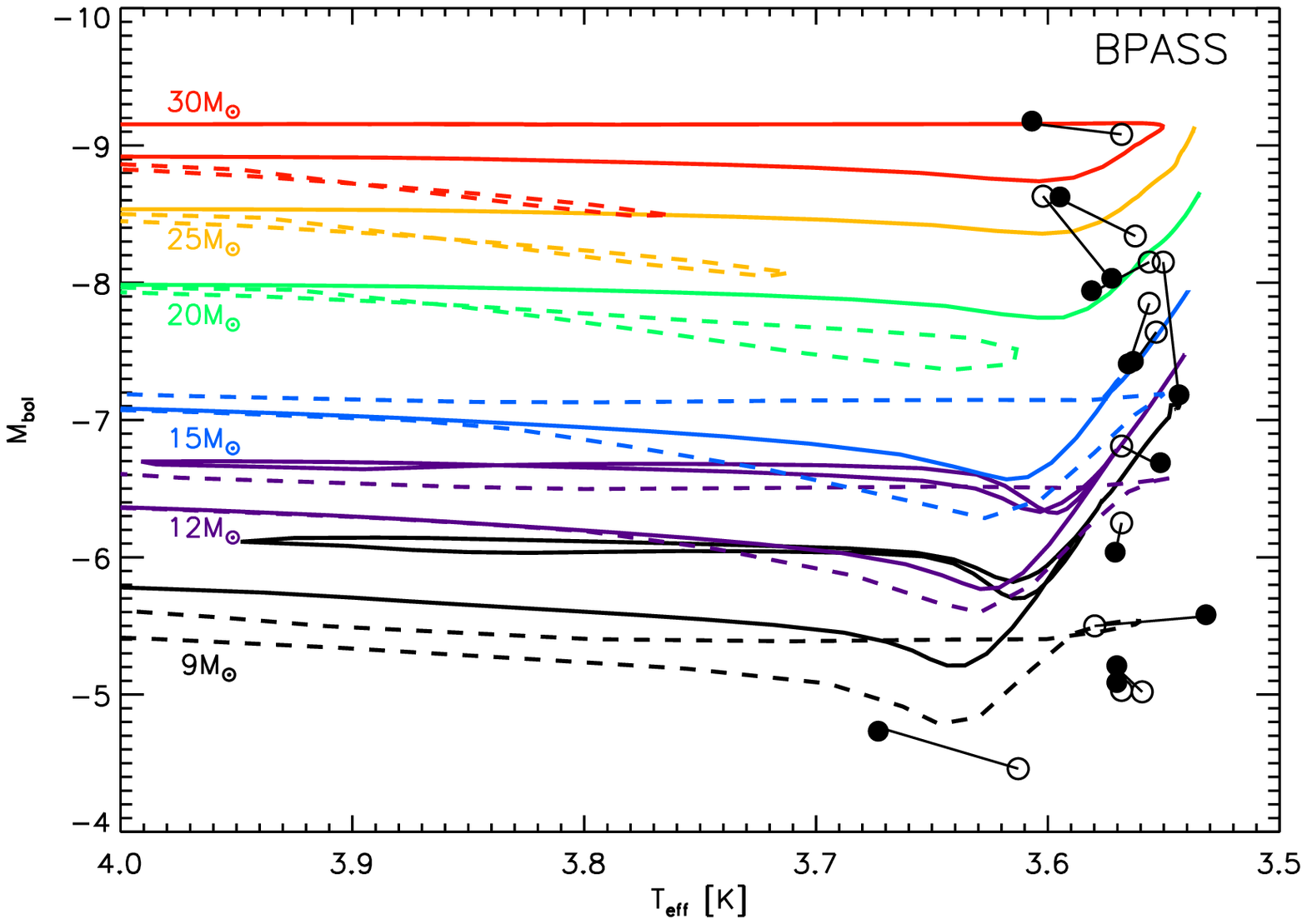}
\caption{Comparing the positions of the sample RSGs on the H-R diagram based on physical properties determined from optical data by L05 (open circles) to those determined from the R09 observations and the near-IR photometric diagnostics derived from the PHOENIX models (filled circles). The near-IR $T_{\rm eff}$ values have been computed from the RSGs' simulated F090W-F150W colors - the best available color index from the R09 data - while the near-IR $M_{\rm bol}$ values are drawn from simulated absolute F090W photometry and the $T_{\rm eff}$-dependent F090W bolometric corrections (adopting the near-IR-based $T_{\rm eff}$ values). The two data points for each star are connected with solid black lines. The RSG positions are compared to two different sets of evolution tracks: the Geneva solar-metallicity non-rotating (solid lines) and rotating (dashed lines) evolutionary tracks from Ekstr\"om et al.\ (2012; top), and the BPASS tracks for single stars (solid lines) and the primary star in a binary with $P=100$ and a secondary-to-primary mass ratio of 0.2 (Eldridge \& Stanway 2016; bottom).}
\end{figure*}

Figure 7 compares the performance of the new $T_{\rm eff}$ diagnostics described above to current methods, plotting the R09 sample of RSGs using two different approaches to determining their physical properties and comparing these results to the predictions of the Geneva (Ekstr\"om et al.\ 2012) and BPASS (Eldridge \& Stanway 2016) stellar evolutionary tracks. The open circles plot RSG positions determined using the optical spectra of L05: $T_{\rm eff}$ was determined from best-fit MARCS models to their optical spectra and TiO band depths, while $M_{\rm bol}$ was based on their absolute $V$ magnitudes and the $T_{\rm eff}$-dependent $BC_V$s computed from the MARCS stellar atmosphere models (adopting $T_{\rm eff}$ from the MARCS optical fitting). The L05 temperatures have errors of $\pm$25 K for the M stars and $\pm$100 K for the K stars due to the larger uncertainties in fitting optical spectra of K supergiants. The filled circles show the same RSGs with positions based on their simulated photometry from the R09 data and the diagnostics presented here. The $T_{\rm eff}$ were calculated using the best-fit polynomial to the (F090W-F150W) color index (the best available color index from the R09 data's wavelength coverage) given in Table 2. The $M_{\rm bol}$ are based on simulated F090W photometry (correcting from apparent to absolute magnitudes based on the distance moduli given in Table 1) and the associated $T_{\rm eff}$-dependent $BC_{\rm F090W}$ (adopting $T_{\rm eff}$ from the (F090W-F150W) color index and computed using the best-fit polynomial given in Table 3. A comparison of the $T_{\rm eff}$ and $M_{\rm bol}$ values determined from the two methods for each star is given in Table 4.

These two independent sets of $T_{\rm eff}$ and $M_{\rm bol}$ diagnostics both produce good agreement between the RSGs and the predictions of the stellar evolutionary tracks. In addition, both the optical spectrophotometry fitting from L05 and the F090W-F150W color index diagnostic presented here determine a similar mean $T_{\rm eff}$ for the RSG sample ($T_{\rm eff, mean}=3715$ K from L05 as compared to the $T_{\rm eff, mean}=3784$ K based on the {\it JWST} diagnostics, with a mean error of 42 K for the L05 $T_{\rm eff}$ values and an error of 3 K from the F090W-F150W color index), and there is no consistent offset between the two methods (one does not produce consistently warmer $T_{\rm eff}$ than the other). However, it is worth noting that the {\it scatter} in the mean $T_{\rm eff}$ is much larger for the {\it JWST} color diagnostic ($\pm$ 326 K, as opposed to $\pm$ 164 K for the L05 fitting) - while some scatter is of course expected in a diverse sample of RSGs with a range of $T_{\rm eff}$, the difference between the two is methods is worth considering. It is likely that this is due in large part to the use of the (F090W-F150W) color index, since as noted in \S3.1 this is a usable but not ideal $T_{\rm eff}$ diagnostic for RSGs. Use of a color index that is more sensitive to $T_{\rm eff}$, such as (F070W-F200W), will likely improve this result.

The L05 and {\it JWST} methods both determine a very similar mean $M_{\rm bol}$ for the RSG sample, with $M_{\rm bol} = -6.99 \pm 1.57$ based on the L05 method and $M_{\rm bol} = -6.86 \pm 1.43$ based on this work. It is also worth noting here that the BCs were calculated using $T_{\rm eff}$ determined from the F090W-F150W index, which did produce a more significant scatter in the sample's mean $T_{\rm eff}$ than the L05 method as noted above. The $M_{\rm bol}$ values plotted in Figure 7 were determined based on F090W photometry and the associated $BC_{\rm F090W}$; however, calculating $M_{\rm bol}$ based on the F150W photometry and $BC_{\rm F150W}$ instead produced essentially identical results, as shown in Table 4.

While these RSGs agree with the predictions of the non-rotating and rotating Geneva tracks (Figure 7, top) and with the single star BPASS tracks (Figure 7, bottom) on the H-R diagram, the stars in this sample do {\it not} show particularly good agreement with the binary evolutionary tracks from BPASS based on either their optical or IR-derived physical properties. Binary evolution predicts lower luminosities and warmer $T_{\rm eff}$ for RSGs, particularly at higher masses ($\ge$20M$_{\odot}$), as a result of mass loss induced by Roche lobe overflow in an interacting binary system. However, as none of these stars shows any signs of binary companions or interactions in their spectra (see Levesque 2017, Neugent et al.\ 2018) this is not a surprising result.

\begin{deluxetable}{l c c c c c c}
\tabletypesize{\scriptsize}
\tablewidth{0pc}
\tablenum{4}
\tablecolumns{7}
\tablecaption{\label{tab:params} RSG Physical Properties from Optical Spectroscopy and {\it JWST} Near-IR Photometry Diagnostics}
\tablehead{
\colhead{Star}
&\multicolumn{2}{c}{$T_{\rm eff}$}
&\colhead{}
&\multicolumn{3}{c}{$M_{\rm bol}$} \\ \cline{2-3} \cline{5-7}
\colhead{}
&\colhead{L05}
&\colhead{This work}
&\colhead{}
&\colhead{L05}
&\multicolumn{2}{c}{This work} \\ \cline{6-7}
\colhead{}
&\colhead{}
&\colhead{}
&\colhead{}
&\colhead{}
&\colhead{F090W}
&\colhead{F150W}
}
\startdata
HD 236697 &3700 &3724 & &$-$6.25 &$-$6.04 &$-$6.02 \\ 
HD 14469 &3575 &3657 & &$-$7.64 &$-$7.43 &$-$-7.41 \\ 
HD 14488 &3550 &3495 & &$-$8.15 &$-$7.19 &$-$-7.16 \\
HD 23475 &3625 &3718 & &$-$5.02 &$-$5.21 &$-$5.19 \\
HD 35601 &3700 &3560 & &$-$6.81 &$-$6.69 &$-$6.67 \\ 
HD 39801 &3650 &3933 & &$-$8.34 &$-$8.62 &$-$8.61 \\
CD $-$31$^{\circ}$ 4916 &3600 &3676 & &$-$7.85 &$-$7.41 &$-$7.39 \\
HD 63302 &4100 &4712 & &$-$4.46 &$-$4.73 &$-$4.76 \\ 
HD 181475 &3700 &3718 & &$-$5.03 &$-$5.09 &$-$5.07 \\
HD 339034 &4000 &3736 & &$-$8.63 &$-$8.03 &$-$8.02 \\ 
BD +39$^{\circ}$ 4208 &3600 &3812 & &$-$8.15 &$-$7.94 &$-$7.93 \\ 
HD 206936 &3700 &4045 & &$-$9.08 &$-$9.18 &$-$9.16 \\ 
HD 216946 &3800 &3402 & &$-$5.50 &$-$5.58 &$-$5.57 
\enddata	      	      	
\end{deluxetable}

\subsection{Surface Gravity}
An additional complication when identifying RSGs through photometry is the problem of foreground contamination. This is especially challenging when studying extragalactic populations, where foreground K and M dwarfs can be mistaken for background RSGs due to their similar colors and incorrect treatments of their distance. This can be avoided by exploiting the effects of surface gravity in these stars. Massey (1998) was able to separate dwarfs and supergiants on a $V-R$ vs. $B-V$ plot; the effects of line blanketing increase at low surface gravities in these stars and are particularly substantial in the $B$ band due to the large number of weak metal lines in that regime, decreasing the $B$ fluxes of the supergiants and separating them from dwarfs on the color-color plot's $y$-axis. For large photometric samples such a method is the best way of avoiding foreground contamination - spectroscopic diagnostics of luminosity class using features such as the Ca II triplet or H$\alpha$ thus far require high-resolution spectra with a high S/N (e.g. Cenarro et al.\ 2001, Jennings \& Levesque 2016). While {\it Gaia} will help with this problem by comprehensively mapping the foreground population of $K$ and $M$ dwarfs down to $G\sim21$ (the DR2 limit for five-parameter astrometric solutions; Brown et al. 2018) this problem will persist in the {\it JWST} era as observations are pushed to fainter and fainter extragalactic populations, capturing samples of RSGs that can be confused with the coolest and lowest-mass (and thus most numerous) $M$ dwarfs that could fall below Gaia's limits.

Figure 8 presents the complete set of color-color diagrams for the {\it JWST}/NIRCam near-IR wide filters, presenting pairs of colors computed from sets of three or four filters. To facilitate comparisons between the different plots, all are presented with the same $y$-axis scale. The plots compare photometry computed from PHOENIX RSG models to photometry computed from PHOENIX red dwarf models. The dwarf models here are from Husser et al.\ (2013), spanning the same range of $T_{\rm eff}$ as the RSGs but with a surface gravity of log($g$) = 4.5. The dwarf models adopt both solar and sub-solar metallicities ($[M/H]=-0.5$ based on the solar abundances given in Asplund et al.\ 2009) to better estimate thin disk and thick disk dwarf populations in the Milky Way respectively, although metallicity has only a minimal effect on the dwarf models' position in color-color space. Where available these plots also include data for the R09 RSGs used in this work as well as the luminosity class V foreground dwarfs from the R09 library with comparable spectral types; however, since this requires coverage in the R09 data across three or four wide bands, only one such comparison is available, for the (F090W-F115W) vs. (F115W-F150W) color-color plot.

RSGs and dwarfs do {\it not} separate across all color-color spaces, and in many cases the diagrams are ineffective at distinguishing between the two populations. Some separations are also quite small ($\lesssim 0.2 mag$) and would be useless with the intrinsic scatter present in a real sample of K and M stars. The (F090W-F115W) vs. (F115W-F150W) plot illustrates this when comparing the predicted position of the RSGs and dwarfs with actual data: while there is good agreement between the observed and predicted populations and their variations with color, the scatter in the observed population is larger and renders this an ineffective diagnostic. Across the board, the RSGs in these model comparisons do appear to extend to redder colors than their dwarf counterparts along both axes; however, this effect is at least partly due to the limitations of the PHOENIX dwarf models used here, as the R09 dwarf data do NOT show this in the (F090W-F115W) color space. In reality dwarfs can evolve to cooler temperatures than RSGs and thus easily appear as redder in these color-color diagrams.

However, several color-color combinations do stand out as potentially effective tools for separating dwarfs and RSGs. The best is the (F070W-F090W) vs. (F090W-F200W) plot, where the separation between dwarf and RSG populations is the largest. The separation is primarily along the (F090W-F200W) axis, with the RSGs appearing redder. This result can be explained by a combination of three effects. First, RSG spectra show stronger absorption from the TiO band at 0.843$\mu$m and the ZrO band at 0.93$\mu$m, both of which contribute to a decreased flux in F090W (particularly for cooler temperatures and redder RSGs, as seen in our data). Second, the dominant source of continuum opacity in both red dwarfs and RSGs is H$^-$, which reaches a local minimum in the F200W wavelength range (the ``H-hump"; Davies et al.\ 2013). High densities correspond to a higher opacity in these stars (e.g. Iglesias \& Rogers 1996, Lamers \& Levesque 2017), which in turn corresponds to a weaker ``H-hump" continuum flux in the denser atmospheres of the high-gravity dwarfs as compared to RSGs. Finally, higher surface gravities in dwarfs relative to RSGs will lead to broader metal absorption features and a net increase in line blanketing effects for dwarfs, which could become more prominent at the F200W wavelengths where the absorption spectrum is primarily comprised of neutral metal absorption lines (as opposed to the more complex molecular features - including TiO, ZrO, VO, CO, and CN - that dominate absorption effects at both shorter and longer wavelengths). \\

\begin{figure*}
\center
\includegraphics[width=4.9cm]{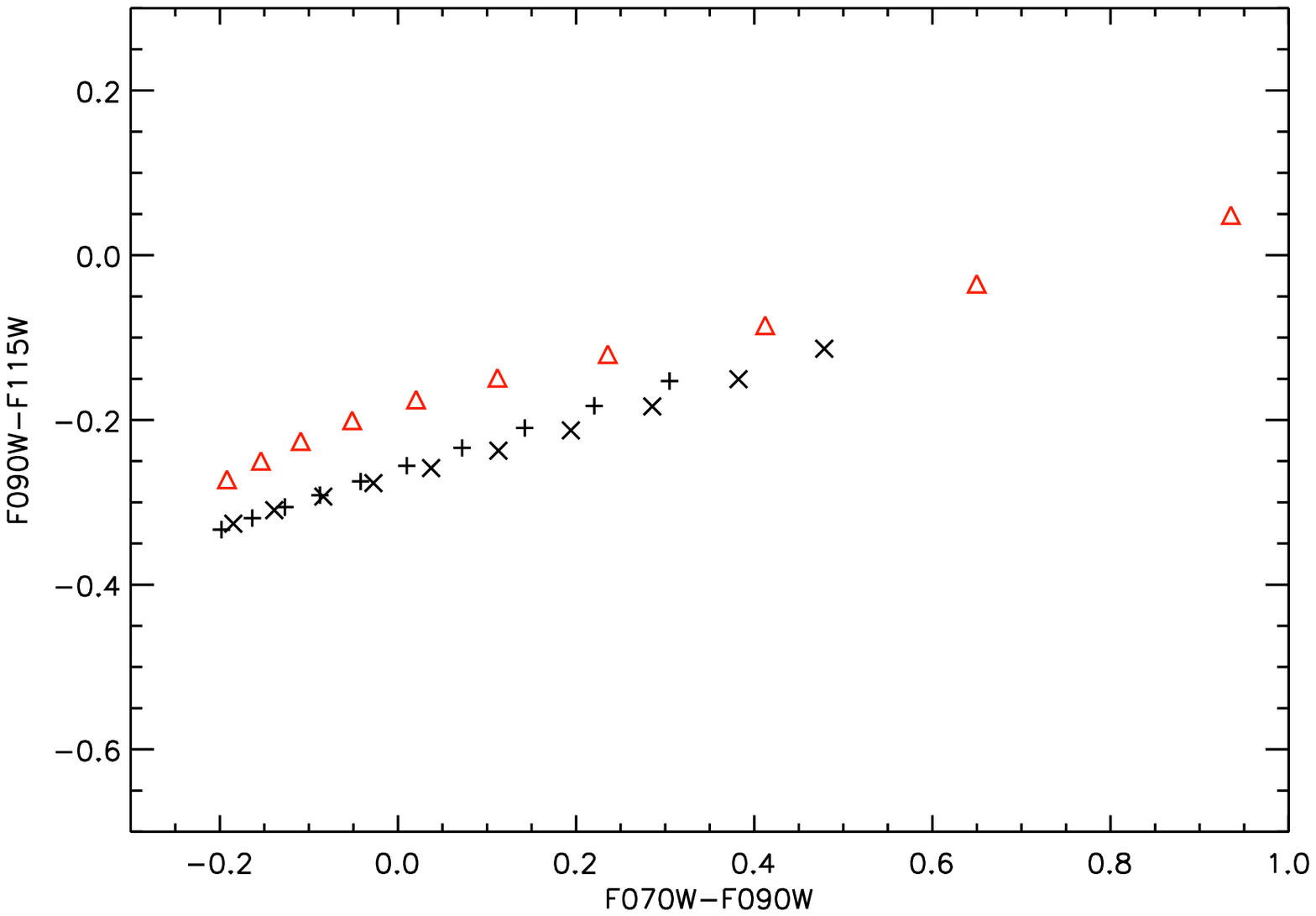} 
\includegraphics[width=4.9cm]{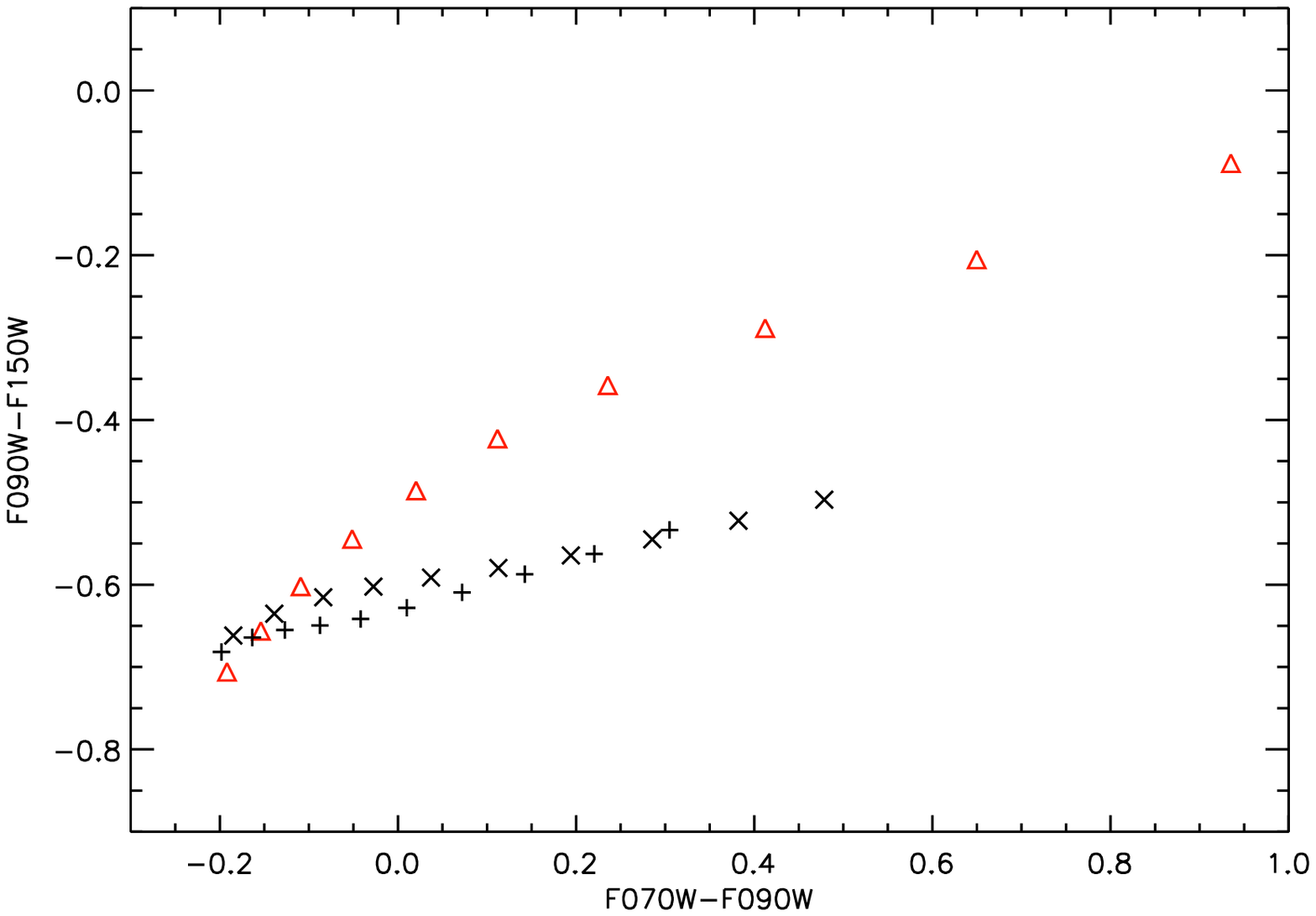} 
\includegraphics[width=4.9cm]{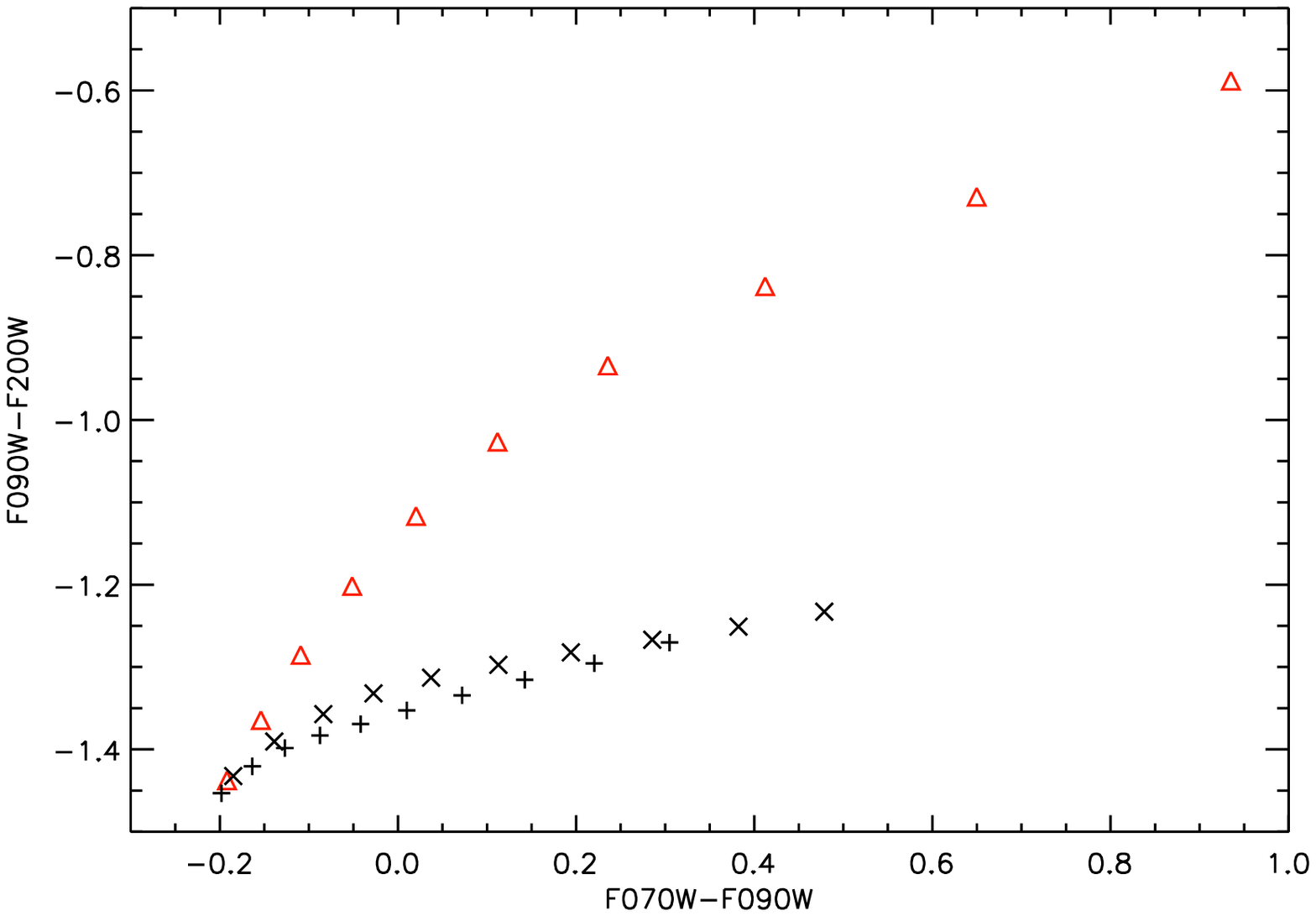} 
\includegraphics[width=4.9cm]{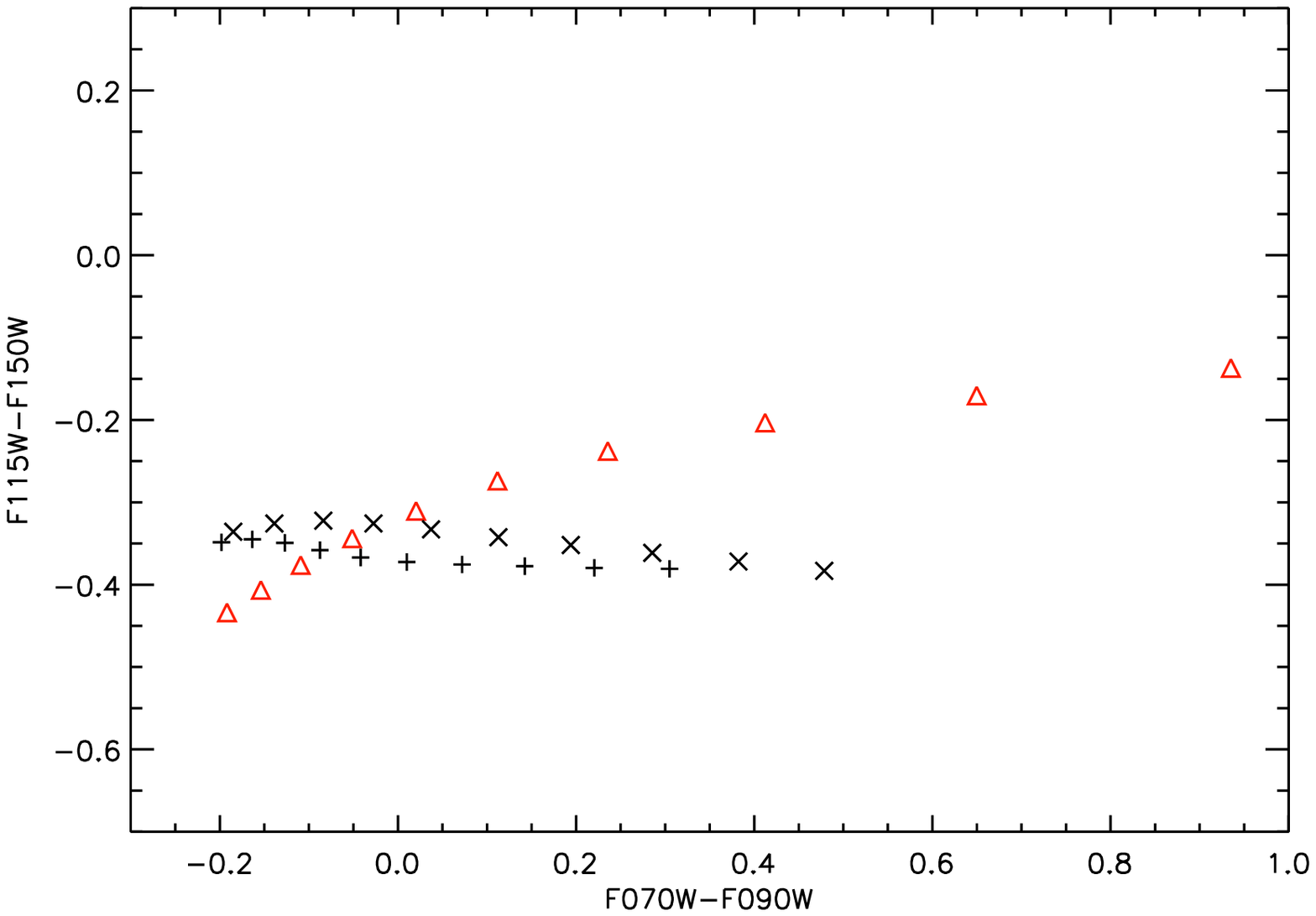} 
\includegraphics[width=4.9cm]{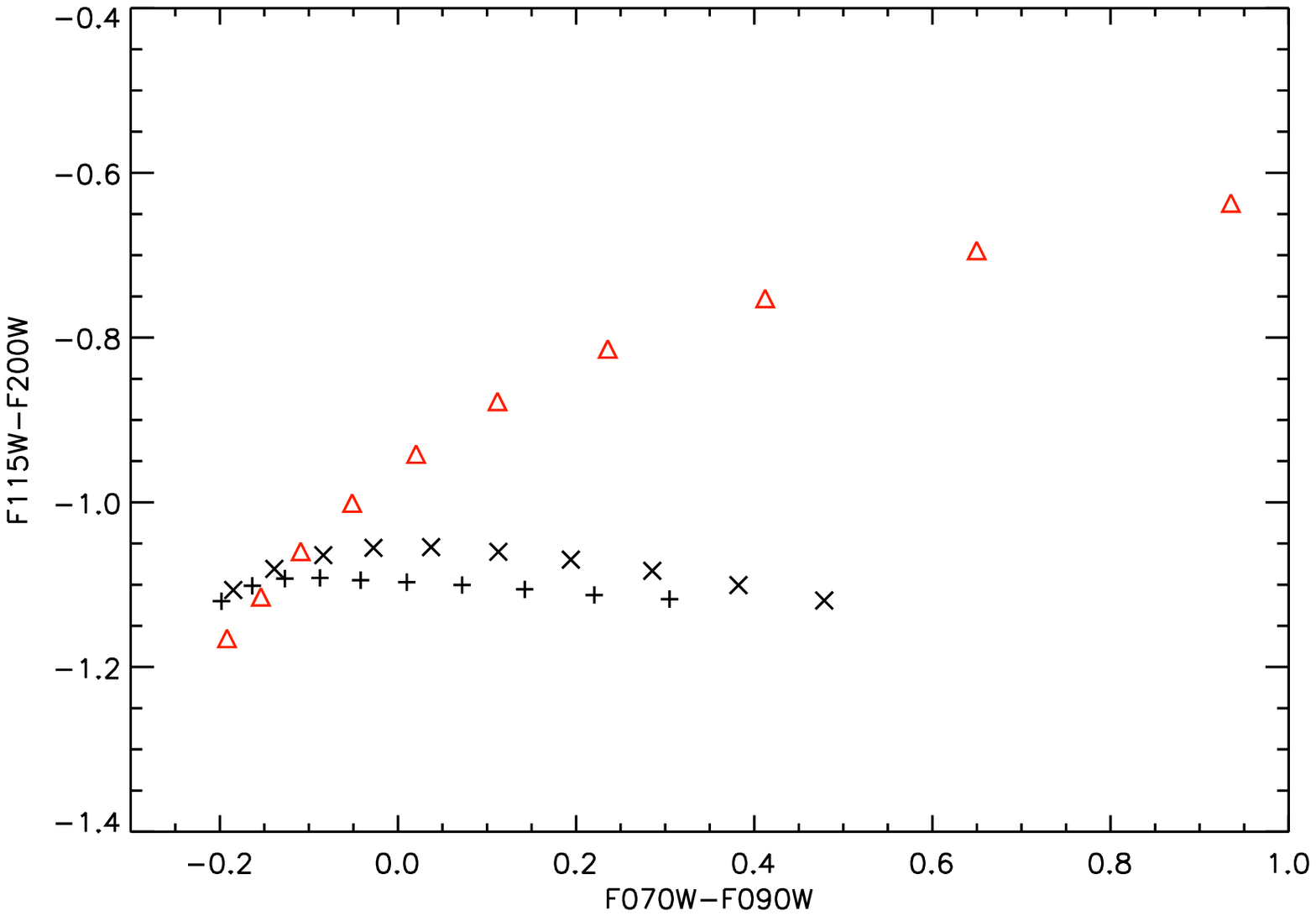} 
\includegraphics[width=4.9cm]{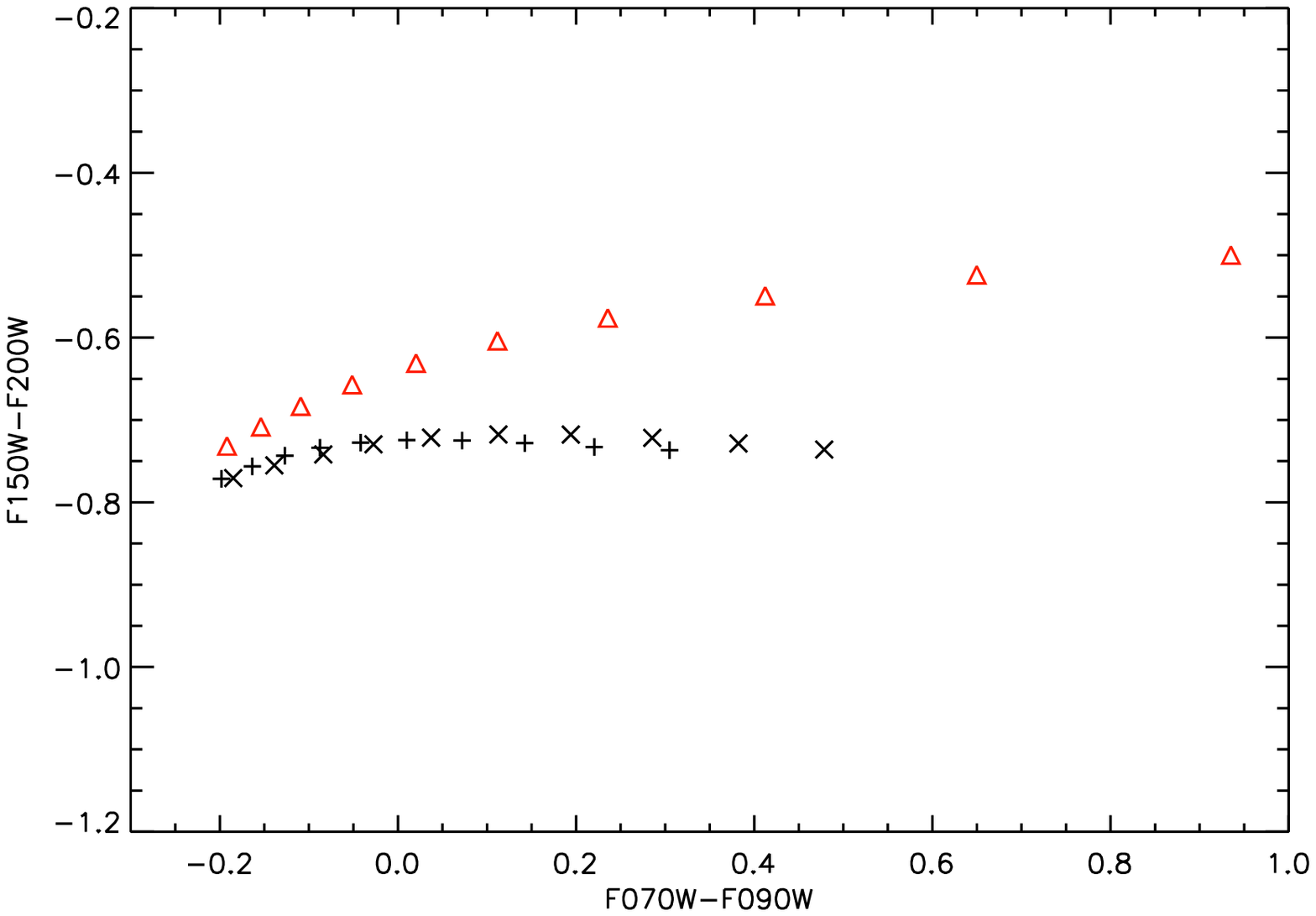} 
\includegraphics[width=4.9cm]{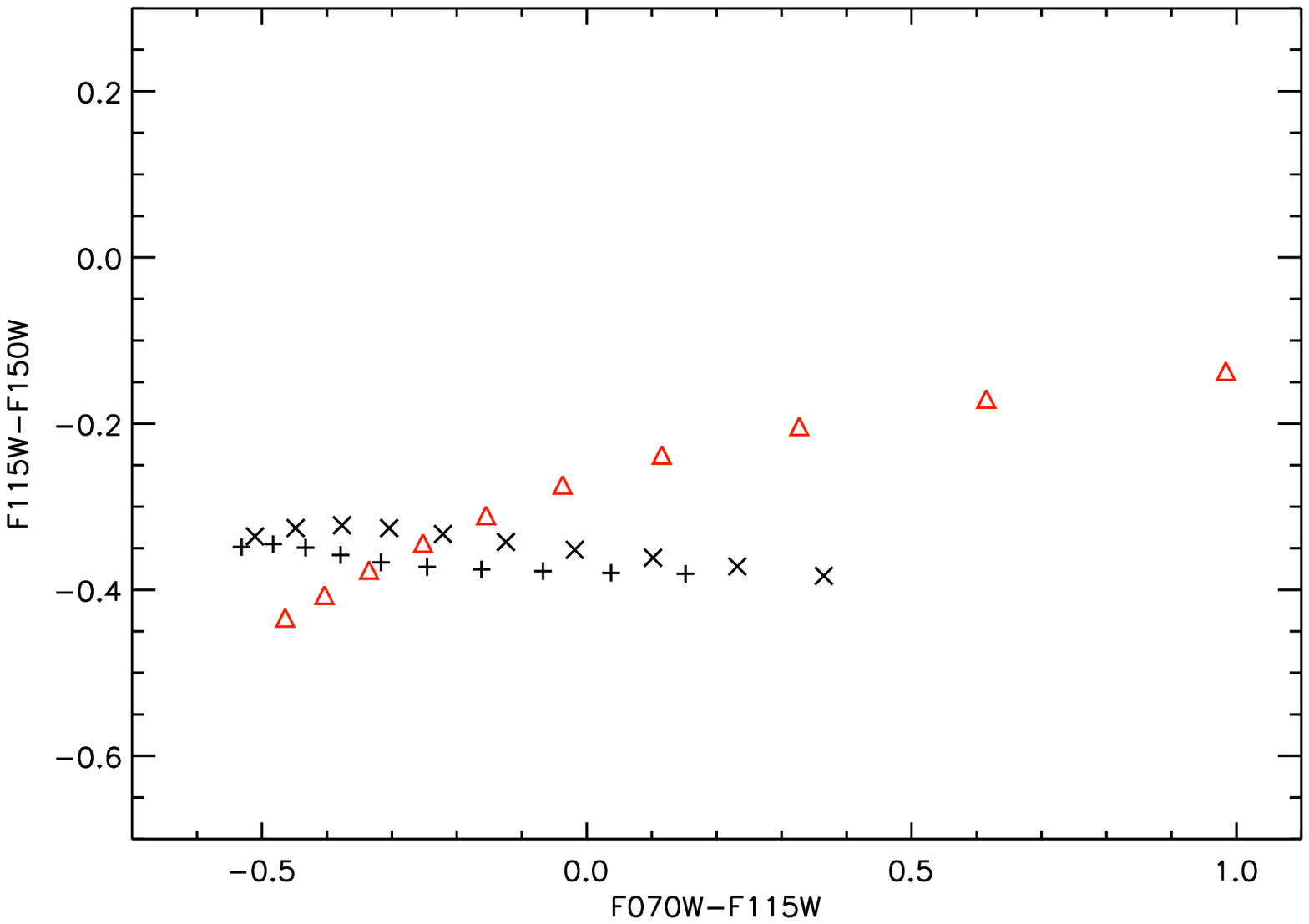} 
\includegraphics[width=4.9cm]{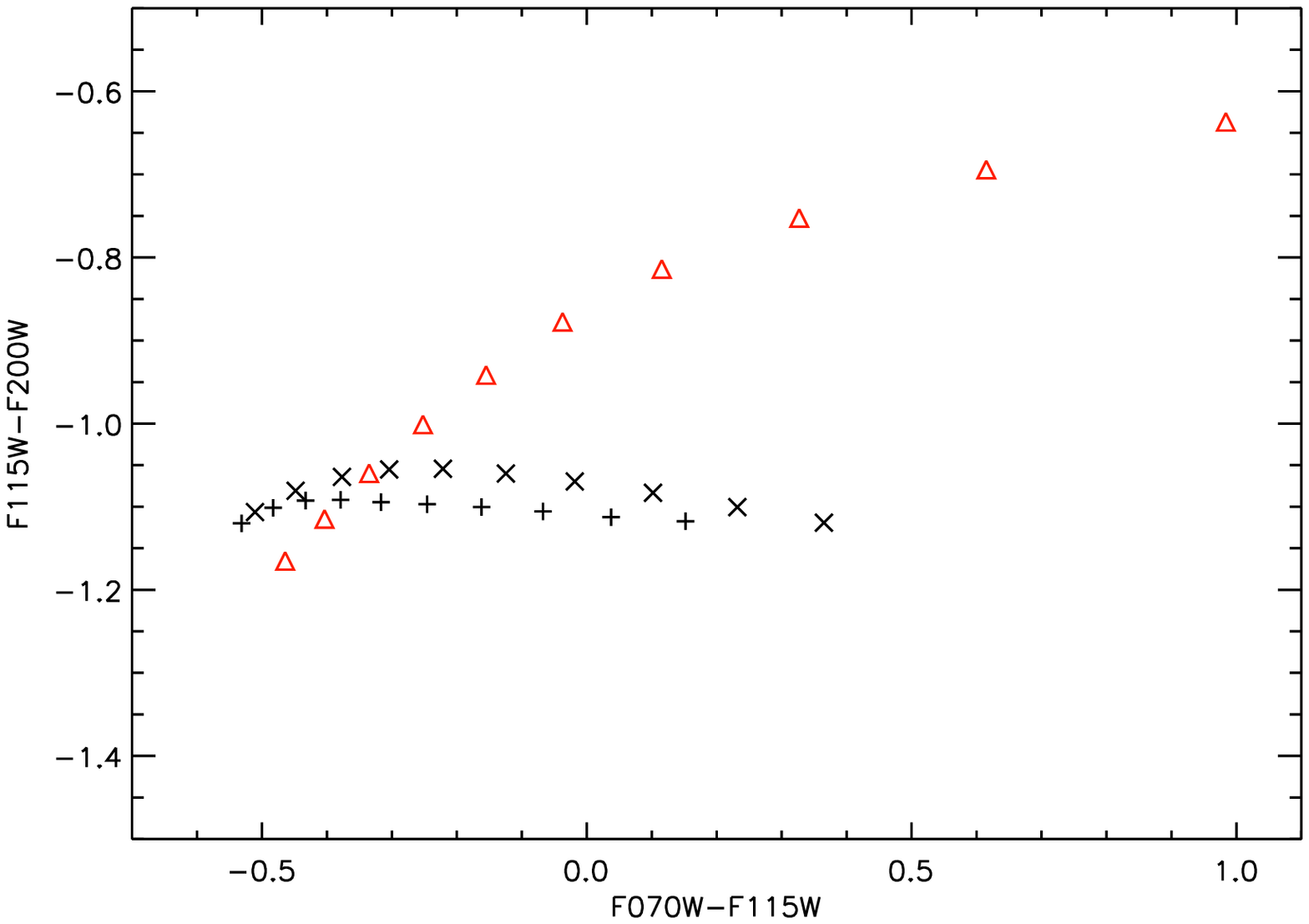} 
\includegraphics[width=4.9cm]{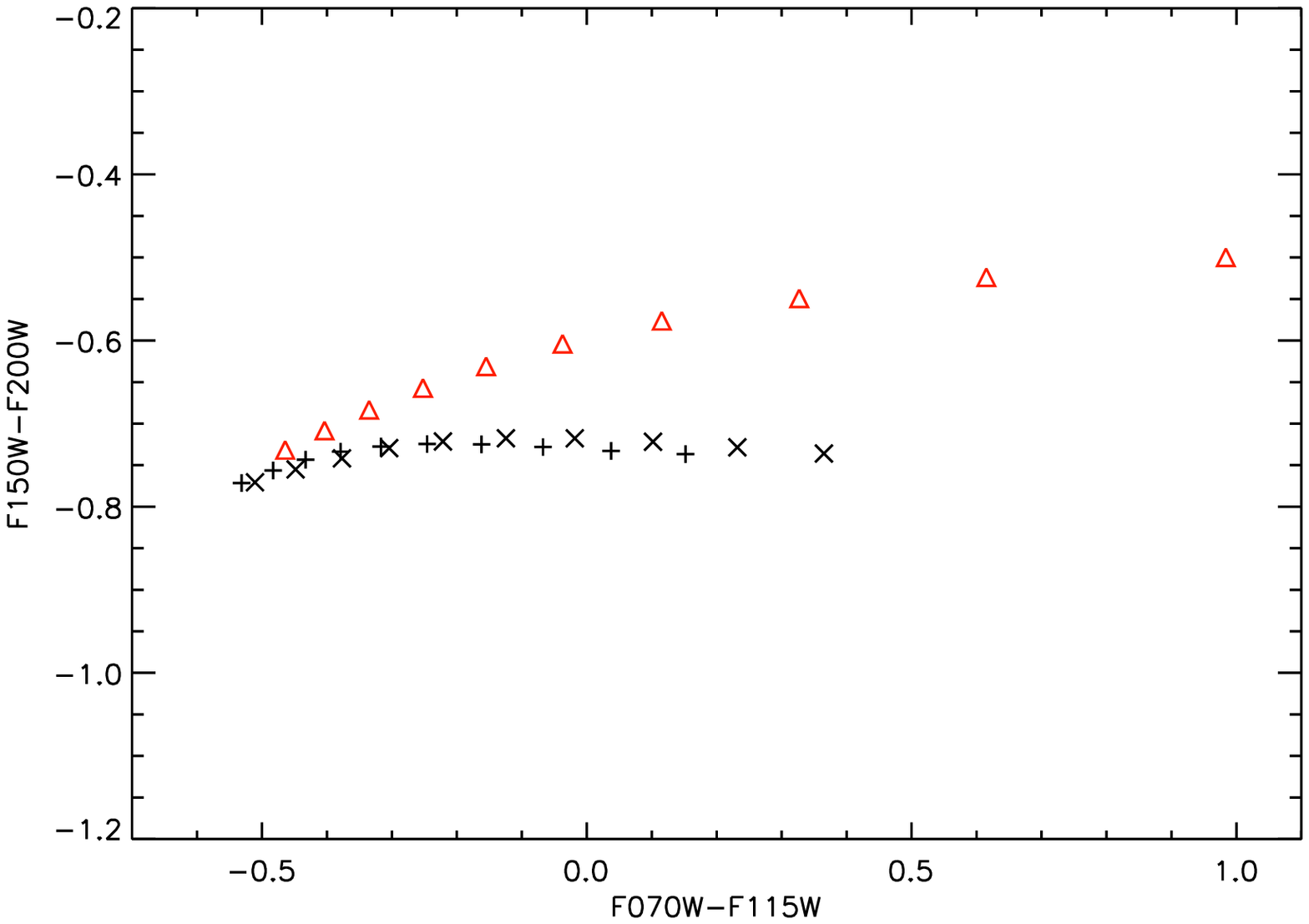}
\includegraphics[width=4.9cm]{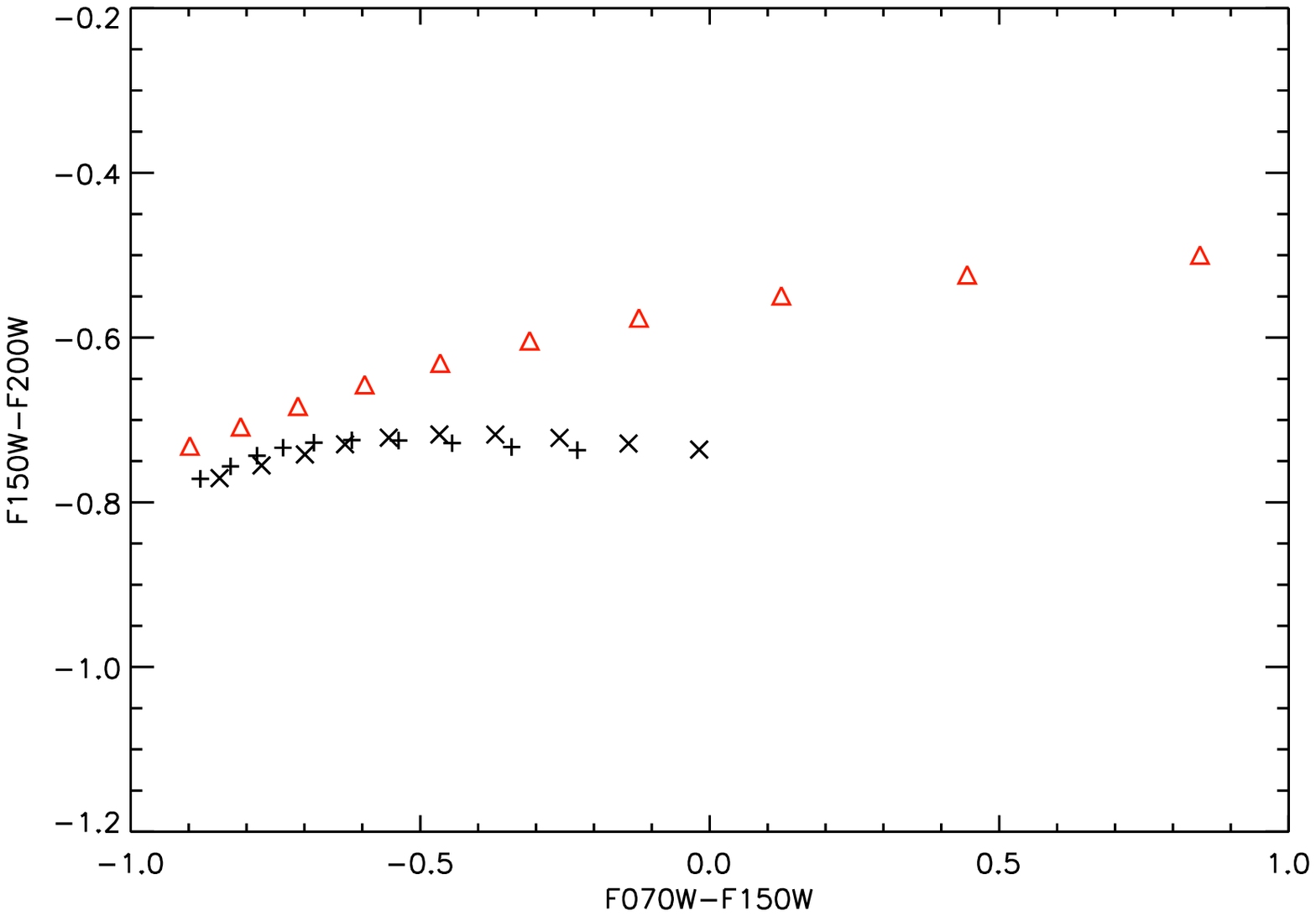} 
\includegraphics[width=4.9cm]{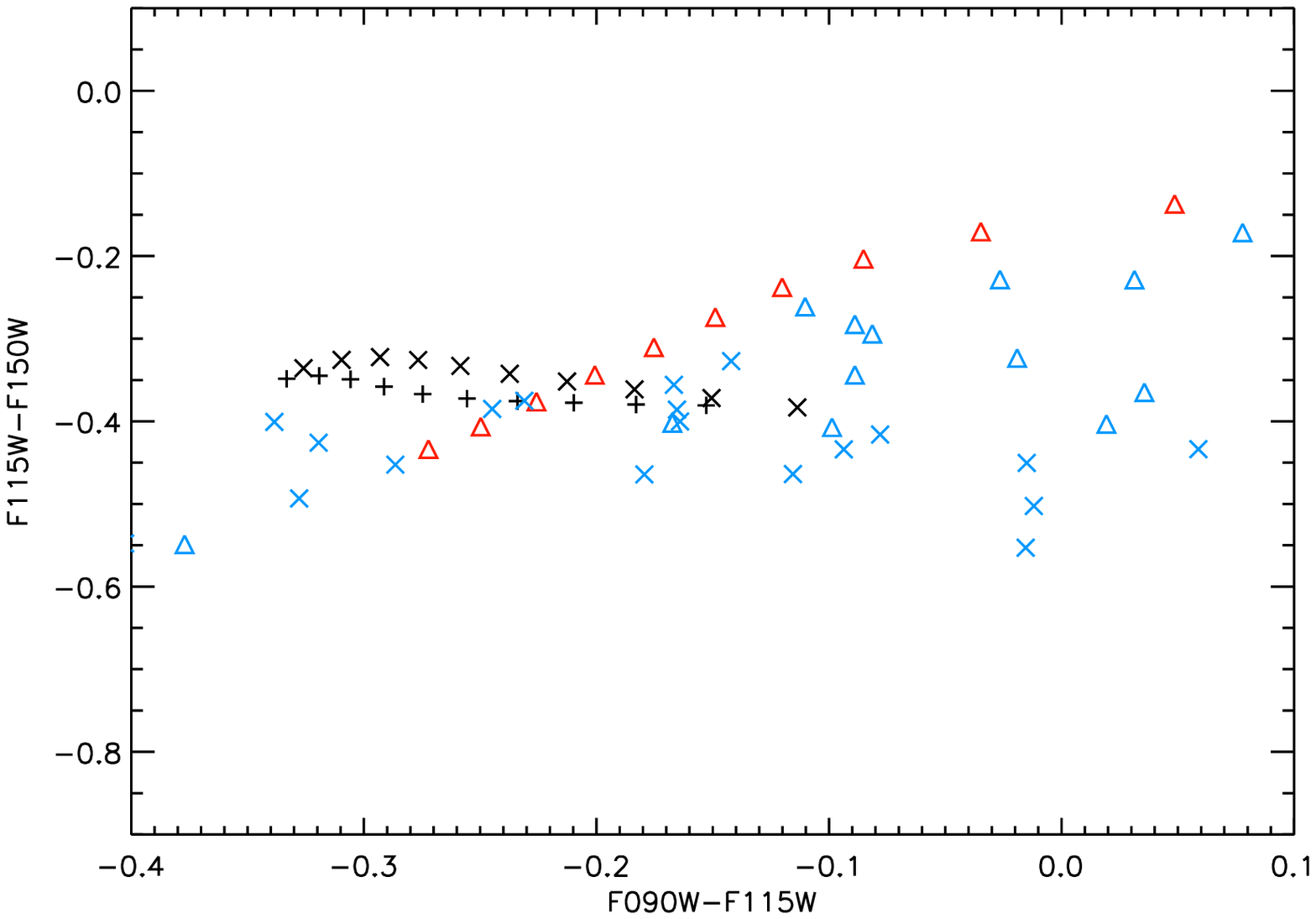}
\includegraphics[width=4.9cm]{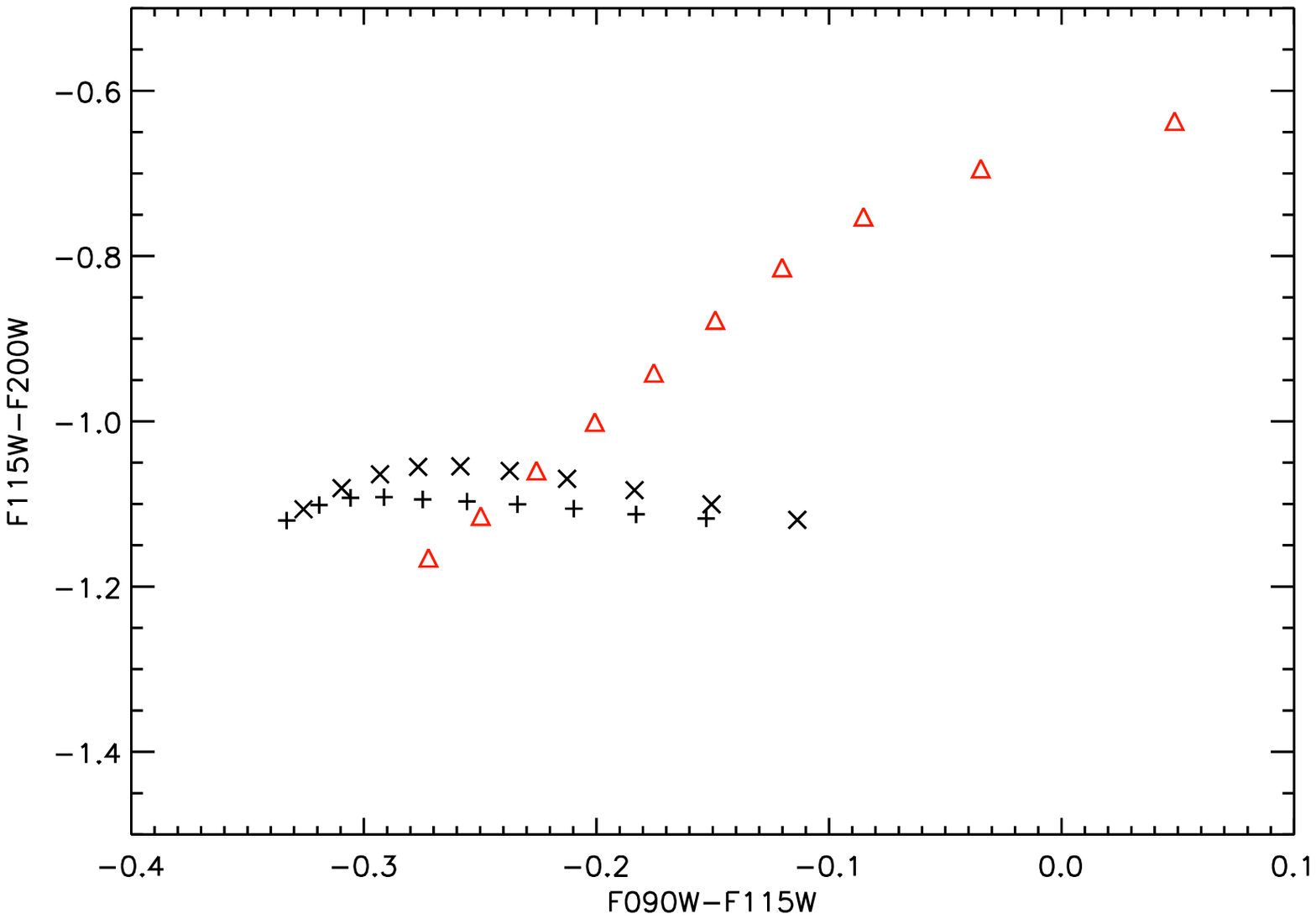} 
\includegraphics[width=4.9cm]{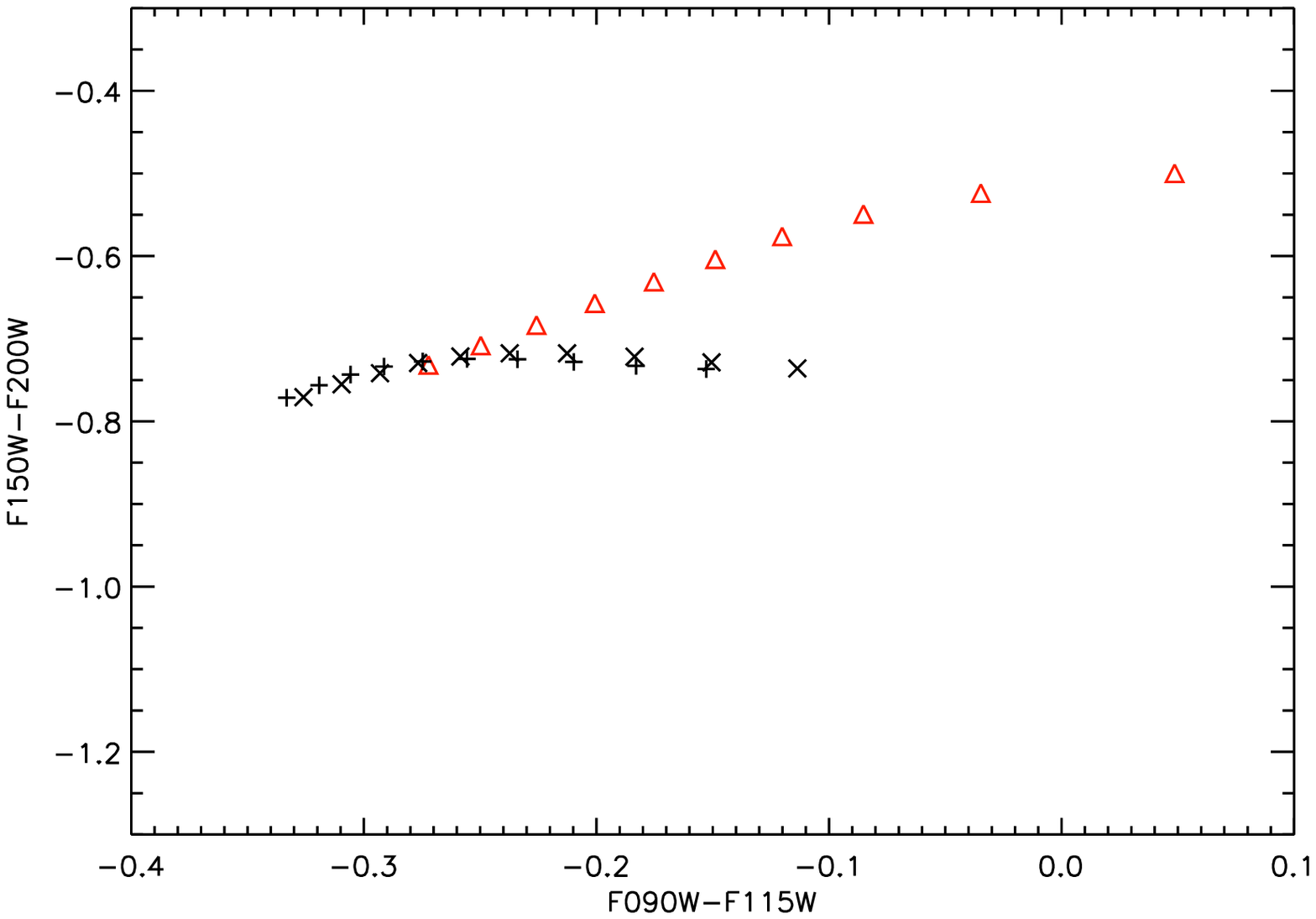} 
\includegraphics[width=4.9cm]{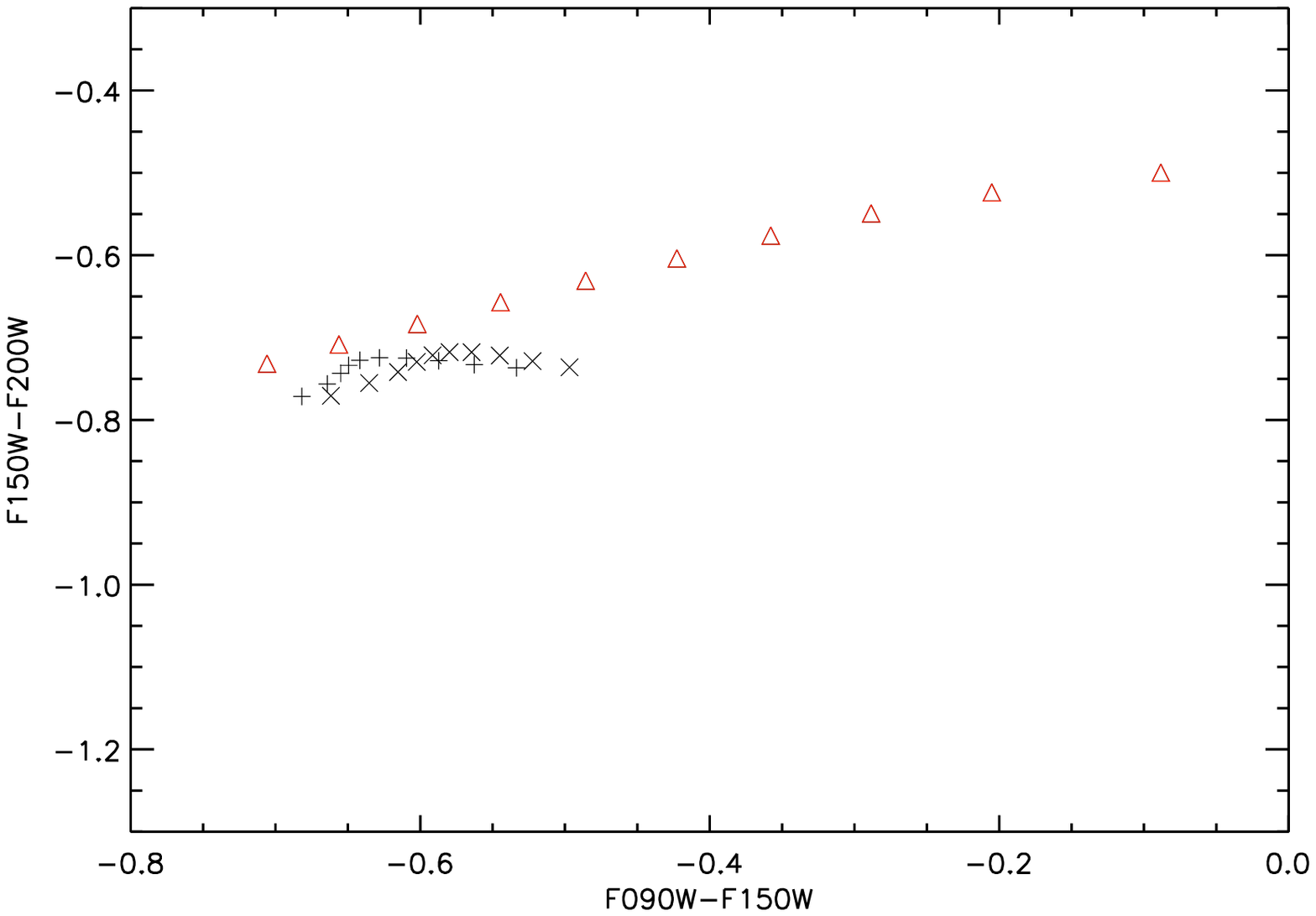} 
\includegraphics[width=4.9cm]{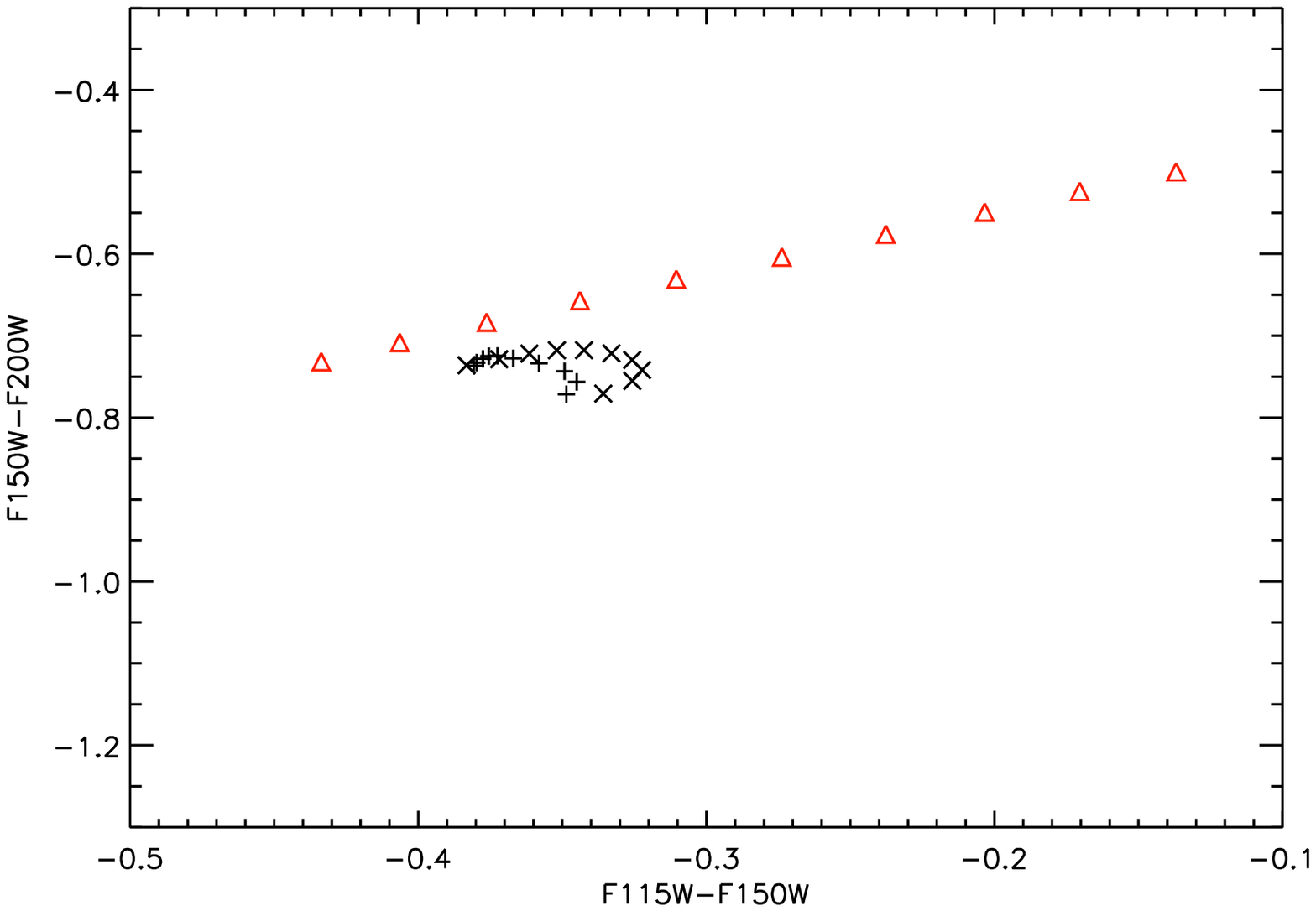} 
\caption{Color-color plots comparing RSG (triangles) and dwarf (crosses) photometry for the NIRCam wide near-IR filters, illustrating the effectiveness of these two-color planes as potential surface gravity diagnostics. Synthetic photometry based on PHOENIX RSG (log($g$)=0.0) spectra are shown in red, while the crosses show photometry based on PHOENIX dwarf (log($g$)=4.5) spectra modeled at both solar metallicity ($\times$'s; [M/H]=0, based on Asplund et al.\ 2009) and sub-solar metallicity (+'s; [M/H] = $-$0.5). In the one color-color plane where the spectral coverage of the R09 library data is sufficient ((F090W-F115W) vs. (F115W-F150)), synthetic photometry from the R09 RSG and dwarf spectra is overplotted in blue. }
\end{figure*}

\section{Discussion and Future Work}
This work has developed photometric tools for determining $T_{\rm eff}$ and $M_{\rm bol}$ in RSGs and for separating RSGs from foreground stars stars using the {\it JWST}/NIRCam near-IR filters. These methods can be immediately applied in the first cycles of {\it JWST} observations to identify RSG populations, place them on the H-R diagram, and determine their physical properties, which in turn will provide us with large new datasets for studying stellar evolution in extragalactic massive star populations and a well-understood sample of candidate core-collapse progenitors.

With observing time at a premium, this work can also be used to identify the handful of NIRCam filters that offer the greatest scientific returns. Future {\it JWST} imaging observations of RSGs should prioritize the F070W, F200W, and F090W filters to amass the most diagnostically-useful set of RSG photometry. F070W-F200W is the most $T_{\rm eff}$-sensitive color index for RSGs, and with the associated BCs we could place stars on the H-R diagram using only these two filters. F090W is the best third filter to add to this list, as it also enables use of the most surface-gravity-sensitive color-color plot (F070W-F090W vs. F090W-F200W) to identify and remove foreground dwarfs from the sample. The F090W BCs are also the smallest and the least sensitive to $T_{\rm eff}$, making it the best filter for determining accurate $M_{\rm bol}$ and minimizing errors in RSGs' vertical position on the H-R diagram (and, as a result, improving estimates of RSG initial masses based on comparisons with stellar evolutionary tracks). The wide filters in general will also be the most efficient choice for extragalactic RSG imaging surveys in particular as they will minimize exposure times. By combining this optimum set of NIRCam filters with observations targeting RSG populations (for example, observing star-forming regions in nearby galaxies with ages that correspond to the RSG evolutionary phase, RSG candidates identified by Spitzer, or a history of hosting multiple Type II-P supernovae that indicates a large population of RSGs and potential core-collapse progenitors, e.g. Smartt 2015, Levesque 2017) we can maximize JWST's already-considerable potential for studying this important stage in massive stellar evolution.

This work has been done for solar metallicity RSGs; extragalactic populations will, of course, cover a broader range of metallicities. Variations in metallicity will affect the strengths of the molecular and atomic absorption features in the RSG spectra as well as the H$^-$ continuum opacity; it is already established that metallicity directly impacts the strengths of the TiO bands in the optical (e.g. Massey \& Olsen 2003, Levesque \& Massey 2012) and the strengths of atomic absorption features in the $J$ band (e.g. Gazak et al.\ 2014). RSG populations as a whole also shift to a warmer mean $T_{\rm eff}$ at lower metallicity as a result of the Hayashi limit's metallicity dependence (e.g. Elias et al.\ 1985, Levesque et al.\ 2006, Drout et al.\ 2012). However, while new calibrations of the above diagnostics will certainly be necessary at different metallicities, the utility of the diagnostics themselves should not be substantially affected; indeed, the separation in (F070W-F200W) color between foreground dwarfs and RSGs should actually increase for lower-metallicity RSGs as a consequence of their decreased H$^-$ opacity.

The role of circumstellar dust is always an important potential complicating factor to keep in mind when studying RSGs. The near-IR is an excellent wavelength regime for studying RSGs along sightlines with large amounts of {\it foreground} dust; previous work has used near-IR colors to identify a large number of RSGs in the inner Milky Way and Galactic Plane, including five RSG-rich clusters (e.g. Figer et al.\ 2006, Davies et al.\ 2007, Clark et al.\ 2009, Negueruela et al. 2010, 2011, Messineo et al.\ 2016). However, the small number of heavily dust-enshrouded RSGs, such as VY CMa, WOH G64, and the other supergiant OH/IR stars may pose a significant challenge even in the near-IR. These stars all show evidence of thick asymmetric circumstellar dust nebulae (e.g. Schuster et al.\ 2006, Ohnaka et al.\ 2008), and observations of these stars have found a larger relative extinction in the IR as compared to the optical, with an inferred circumstellar dust grain size of $\sim$0.5$\mu$m (50 times larger than that of the diffuse ISM; Scicluna et al.\ 2015). Without a clearer understanding of RSG circumstellar dust and its effects, OH/IR RSGs are poor candidates for study with near-IR photometric diagnostics.

The near-IR also represents only a fraction of {\it JWST}'s abilities. Mid-IR photometry has already proven invaluable at identifying RSGs (and is particularly effective at identifying dusty RSGs, which show excess luminosity in the mid-IR). Britavskiy et al.\ (2014) and Messineo et al.\ (2012) presented photometric tools for identifing RSG populations using {\it Spitzer}/IRAC [3.6], [4.5], and [8.0] photometry, and mid-IR colors can also be used as measures of mass loss rates (e.g. Davies et al.\ 2007). Jones et al.\ (2017) also recently identified {\it JWST}/MIRI color indices and classifications that can be used to study a broad variety of dusty stars in the Local Volume, including RSGs. Future work will compute additional RSG diagnostics with the mid-IR (long wavelength channel) {\it JWST}/NIRCam filters that complement the diagnostics described here and in Jones et al.\ (2017), thus offering additional tools for identifying and observing extragalactic RSG populations using JWST.

Future work on this topic will also delve into potential spectroscopic diagnostics available in the near- and mid-IR that can be utilized with future {\it JWST}/NIRSpec observations of RSGs. Gazak et al.\ (2014) have already identified atomic absorption features of Fe I, Ti I, Si I, and Mg I in the $J$-band that can be used as metallicity diagnostics, a valuable tool for measuring stellar (as opposed to gas phase) metallicities in other galaxies. At the moderate-to-low resolution of NIRSpec ($100\lesssim R\lesssim2700$) the broad molecular bands will also be a useful avenue to explore. The bluest reaches of NIRSpec sample the $T_{\rm eff}$-sensitive TiO bands, while the near- and mid-IR ZrO, OH, and SiO molecular absorption bands and the ``H-hump" are all potential $T_{\rm eff}$ diagnostics, and the CN, CO, and SiO bands all show potential sensitivity to surface gravity (see R09). Exploring these spectroscopic RSG diagnostics in the near- and mid-IR would also be more effective with a more comprehensive dataset of observed RSG spectra, ideally with continuous wavelength coverage from the optical to the mid-IR that would make it possible to compare new diagnostics with existing methods. To establish a solid theoretical framework for studying RSGs, future work will also include a detailed comparison of current stellar atmosphere models for RSGs in order to quantify their utility and limitations in both the optical and IR. This will make it possible to identify the best models to use in conjunction with {\it JWST} observations, and to make further improvements to the next generation of stellar atmosphere models for RSGs. \\

\acknowledgements
The author thanks Rubab Khan, Jamie Lomax, Philip Massey, Kathryn Neugent, Bertrand Plez, and Sarah Tuttle for using discussions regarding this work, as well as the anonymous referee for their helpful and constructive comments. This research was supported by a fellowship from the Alfred P. Sloan Foundation.

\end{document}